\documentclass[twocolumn,prd,nofootinbib,aps,floats,floatfix,amsmath,amssymb,secnumarabic]{revtex4-1} %
\usepackage[final]{graphicx}
\usepackage{amsmath}
\usepackage{bbm}
\usepackage{amsfonts}
\usepackage{amssymb}
\usepackage{latexsym}
\usepackage{graphicx}
\usepackage[english]{babel}
\usepackage{multirow}
\usepackage{float}
\usepackage{url}
\usepackage{hyperref}
\usepackage{slashed}
\usepackage{xcolor}
\usepackage{verbatim} 
\usepackage[utf8]{inputenc}

\def\sfrac#1#2{{\textstyle{#1\over #2}}}
\def\HH{\mathrm{H}}
\newcommand{\be}{\begin{equation}}
\newcommand{\ee}{\end{equation}}
\newcommand{\ba}{\begin{array}}
\newcommand{\ea}{\end{array}}
\newcommand{\bea}{\begin{eqnarray}}
\newcommand{\eea}{\end{eqnarray}}
\newcommand{\sss}{\scriptscriptstyle}

\newcommand{\de}{{\mathbf e} }
\newcommand{\pd}{{\mathbf p} }
\newcommand{\dH}{{\mathbf H} }

\newcommand{\tr}{{\rm tr}}
\newcommand{\Tr}{{\rm tr}}

\begin{document}

\title{Composite strongly interacting dark matter}
\author{James M.\ Cline}
\author{Zuowei Liu}
\author{Guy D.\ Moore}
\affiliation{Department of Physics, McGill University,
3600 Rue University, Montr\'eal, Qu\'ebec, Canada H3A 2T8}
\author{Wei Xue}
\affiliation{INFN, Sezione di Trieste, SISSA, via Bonomea 265, 
34136 Trieste, Italy}

\begin{abstract}
It has been suggested that cold dark matter (CDM) has
difficulties in explaining tentative evidence for noncuspy halo profiles
in small galaxies, and the low velocity dispersions observed in the
largest  Milky Way satellites (``too big to fail'' problem).  Strongly
self-interacting dark matter has been noted as a robust
solution to these problems.  The elastic cross sections required are much larger
than predicted by generic CDM models, but could naturally be of the right
size if dark matter is composite.  We explore in a general way the constraints
on models where strongly interacting CDM is in the form of 
 dark ``atoms'' or ``molecules,'' or bound states of a
confining gauge interaction (``hadrons'').  
These constraints include
considerations of relic density, direct detection, big bang nucleosynthesis, the
cosmic microwave background, and LHC data.
\end{abstract}

\maketitle

\section{Introduction} 
\label{intro} Cold dark matter (CDM) has proven in most respects
to be an excellent description of the 22\% of the universe's energy
density that is not baryonic nor dark energy.  There are however a 
few suggested problems in its ability to predict some of the observed
properties of dark matter halos.  These are the behavior of the
density profile near galactic centers, which comes out too cuspy in
$N$-body simulations 
\cite{Borriello:2000rv,Donato:2009ab,deBlok:2009sp,deBlok:2002tg}, 
as well as the 
overabundance of prominent
satellite galaxies, relative to observations: the ``too big to fail''
(TBTF) problem \cite{BoylanKolchin:2011de,BoylanKolchin:2011dk}.  Long ago it was
pointed out that strong elastic scattering of dark matter with itself,
with cross section per mass $\sigma/m \sim 0.1-1$ cm$^2$/g,\footnote{
This is related to the alternate unit of cross section per mass
1 b/GeV $=0.56$ cm$^2$/g, where 
b $=$ barn $= 10^{-28}$m$^2$.  We take
$c=1$ throughout.}
would ameliorate the first of these problems 
\cite{Spergel:1999mh,Dave:2000ar}.  More recent work has shown that
the TBTF problem can also be addressed in this way
(see ref.\ \cite{Weinberg:2013aya} for a review).\footnote{A recent paper
\cite{Polisensky:2013ppa} finds that the TBTF problem is ameliorated by updating the
values of cosmological parameters that go into the simulations.}

The idea of strongly interacting dark matter (SIDM) fell out of favor
in light of subsequent arguments that $\sigma/m$ should be less than
$0.02$ cm$^2$/g to avoid making observed elliptical halos become too
spherical \cite{MiraldaEscude:2000qt}.   A similar but weaker upper 
limit of 0.7 cm$^2$/g was found using simulations of the Bullet Cluster 
\cite{Randall:2007ph}.  The arguments leading to the more stringent
bound have recently been reexamined \cite{Peter:2012jh} in light of
improved simulations, leading the authors to conclude that the halo
ellipticity bound should be relaxed to the level of 0.1 cm$^2$/g.  The
same authors argue that this value is moreover
consistent with what is needed to solve the problems of halo cuspiness
and excess substructure \cite{Rocha:2012jg}.  Subsequently ref.\
\cite{Zavala:2012us} studied this issue using a higher resolution
simulation and concluded that a larger value of 0.6 cm$^2$/g $=1.1$
b/GeV is needed
to produce the cores inferred in dwarf galaxies by ref.\ 
\cite{Walker:2011zu}.\footnote{Ref.\ \cite{Strigari:2010un} shows that
different assumptions about the parametrization of the dwarf halos
than made in \cite{Walker:2011zu} can significantly reduce the 
evidence for cores in these systems, though not disprove their
existence.}  We adopt this figure in the following for the preferred
value of the SIDM cross section.\footnote{This value of $\sigma/m$ may
start to be in marginal conflict from halo ellipticity bounds, which
limits  $\sigma/m < 1$ cm$^2$/g \cite{Peter:2012jh}.  More detailed 
investigations using numerical simulations of halo shapes for 
intermediate values of $\sigma/m$ will be needed to settle
the question.}

To appreciate the challenge of achieving such a large cross section if
dark matter is a fundamental particle, consider scalar DM
with a quartic interaction $(\lambda/4!) S^4$.  The cross section over
mass is given by $\sigma/m = \lambda^2/(128\pi) (m/{\rm GeV})^{-3}$
$\cdot\, 4\times  10^{-4}\,$b/GeV.  Even at the largest sensible value
of $\lambda\sim 32\pi^2/3$, where the one-loop correction 
to $\sigma$ becomes of the same order as the tree level cross section,
to reach the
level of $\sigma/m = 0.1$ cm$^2$/g requires a small
dark matter mass, $m \sim 400$ MeV, introducing a new hierarchy
problem worse than that of the weak scale. 
  It is possible to overcome this
limitation in a more complicated model where heavy dark matter
interacts with itself via a light vector boson, with mass $m_V\lesssim
1-100$ MeV \cite{Feng:2009hw}-\cite{Kaplinghat:2013kqa}.  
But here the question of naturalness
has just been transferred to the vector boson mass scale (except
in the limit where it is massless \cite{Ackerman:mha}).

On the other hand, normal atoms and nuclei in the visible sector
have $\sigma/m$ close to or above the values of interest.
The large cross section of atoms arises because they
themselves are large, due to being weakly bound.  For nuclei the cross
section is large because of the residual strong interactions, mediated
by relatively light mesonic or nuclear bound states.  It is therefore
interesting to consider dark analogues of these kinds of states in a
hidden sector as candidates for dark matter.  Models of
atomic dark matter have been previously considered (although starting
from different motivations) in refs.\ 
\cite{Kaplan:2009de}-\cite{Cline:2013pca}.%
\footnote{We do not
consider the scenario of $^4$He X$^{--}$ bound states of ref.\ 
\cite{Khlopov:2011tn} since exotic stable 
X$^{--}$ particles appear to
be ruled out by big bang nucleosynthesis constraints \cite{Jedamzik:2009uy}, which
are far stronger for charge-2 relics than the usually studied charge-1
relics \cite{maxim}, and by
anomalous
hydrogen constraints \cite{Yamagata:1993jq}.
For directly detecting such bound state dark matter, see e.g.\ \cite{Laha:2013gva}. 
}
Historically, the first atomic dark matter model was in the context
of mirror symmetry, in which the dark sector is an exact copy of the
visible one (see refs.\ 
\cite{Berezhiani:2003xm}-\cite{Foot:2004pa} for a review).  We do not
consider this scenario here, since we will show that the dark electron
is always much heavier than $m_e$ in the models that give
the desired self-interaction cross sections.  The case in which mirror
symmetry is broken \cite{Berezhiani:2008zza} might at first seem to offer a greater possibility to provide a concrete
realization of this scenario, but we will show in section 
\ref{sec:adm} that it is also incompatible with our criteria.

Composite (``hadronic'') dark matter models involving confining gauge forces have
been considered in refs.\
\cite{Faraggi:2000pv}-\cite{Holthausen:2013ota}, 
with much of the recent motivation stemming from observations of DAMA
\cite{Bernabei:2010mq} and other direct detection experiments, or the
idea of linking dark matter genesis to baryogensis and thus explaining
their similar abundances.  (Indeed a common attribute of 
atomic and ``baryonic'' DM models is that they are asymmetric, with the
relic abundance arising analogously to the baryon asymmetry rather
than by thermal freeze-out.)  Here we add to the previously considered
motivations by emphasizing the natural capacity of composite dark
matter for having strong enough self-interactions to overcome the halo
structure problems. 

The main particle physics alternative to SIDM for addressing the shortcomings of
CDM has been warm dark matter (WDM), with mass of order keV; see ref.\ 
\cite{Biermann:2013nxa} for a recent review.\footnote{Ref.\ \cite{Marsh:2013ywa} recently proposed that ultra-light axions
comprising 85\% of the total dark matter could provide an alternative
solution to the problems of CDM.}\ \   Ref.\ 
\cite{Weinberg:2013aya} argues that warm dark matter of a given mass
is not able to solve the halo structure problems
while remaining consistent with Lyman-$\alpha$ determinations of the power
of density fluctuations on small scales 
\cite{Seljak:2006qw}-\cite{Schneider:2013wwa}.  The latter place a lower
limit of at least 4 keV on the dark matter mass, which is too large
to allow for effective smoothing of central cusps of galactic halos.
It should be noted however that this conclusion depends upon the
assumed value of the Milky Way halo mass $M_{\rm halo}$; if 
$M_{\rm halo} > 1.4\times 10^{12}\,M_\odot$, then lower WDM masses can be
tolerated \cite{Kennedy:2013uta}.

In section \ref{sec:adm}  we outline the requirements of
atomic DM models to have a strong enough
self-interaction cross section.  Here we also treat the possibility
that the dark matter is primarily in molecular form, finding a larger
region of viability to be SIDM.  We discuss constraints from direct
detection and cosmology on the atomic models.
In section \ref{sec:meson} we turn to the possibility of dark
``mesons'' in a strongly coupled dark sector, showing that they can be
SIDM if sufficiently light (30-100 MeV).  To get the right relic
density by thermal production, we argue that the hidden quarks 
should interact with massless dark photons that kinetically mix
with the normal photon, and we demonstrate an explicit model,
discussing the cosmological constraints that apply.
The case of hidden sector ``baryons'' as the dark matter  is
examined in section \ref{sec:baryon}, and that of glueballs in 
section \ref{gball}.  We summarize our results in section
\ref{conclusions}.

\section{Atomic dark matter}
\label{sec:adm}

We first examine the simplest example of atomic dark matter
\cite{Kaplan:2009de}, a bound state of elementary particles
transforming under a hidden U(1)$'$ symmetry with charge $g'$.  The
constituents are the dark ``proton''  $\pd$ and ``electron'' $\de$,
assumed to be spin-1/2 particles.  The analogues of the fine
structure  constant and Bohr radius are $\alpha' = g'^2/4\pi$ and
$a_0'= (\alpha'\mu_\dH)^{-1}$ respectively, where $\mu_\dH = m_\de
m_\pd/(m_\de+m_\pd)$ is the reduced mass.  Taking account of binding
energy $E_b \cong \alpha'^2\mu_\dH/2$, the mass of the ground state dark
atom is $m_\dH = m_\pd+m_\de -E_b$.  We will also introduce the
mass ratio $R \equiv m_{\pd}/m_{\de}$, which should be treated as
a model parameter.  It enters in the scattering cross section through
the combination
\be
	{m_\dH\over \mu_\dH } \cong R + 2 + R^{-1}\equiv f(R)
\label{fRrel}
\ee
where we have ignored the binding energy contribution to $m_\dH$.
It has  been shown in
\cite{Kaplan:2009de,CyrRacine:2012fz} that as long as $\alpha'$ is
sufficiently large ($\gtrsim 10^{-2}$, as we will verify for most of
the relevant parameter space), the ionized
fraction of the atoms is suppressed, and dissipative processes that
would lead to collapse of the halo and formation of dark stars are
negligible.  Thus dark halos will not differ radically relative to
expectations for CDM.  Only those more subtle properties that we want to
alter will be affected by the strong elastic self-interactions.

\begin{figure*}[t]
\vskip-0.75cm
\centerline{\includegraphics[width=1.1\columnwidth]{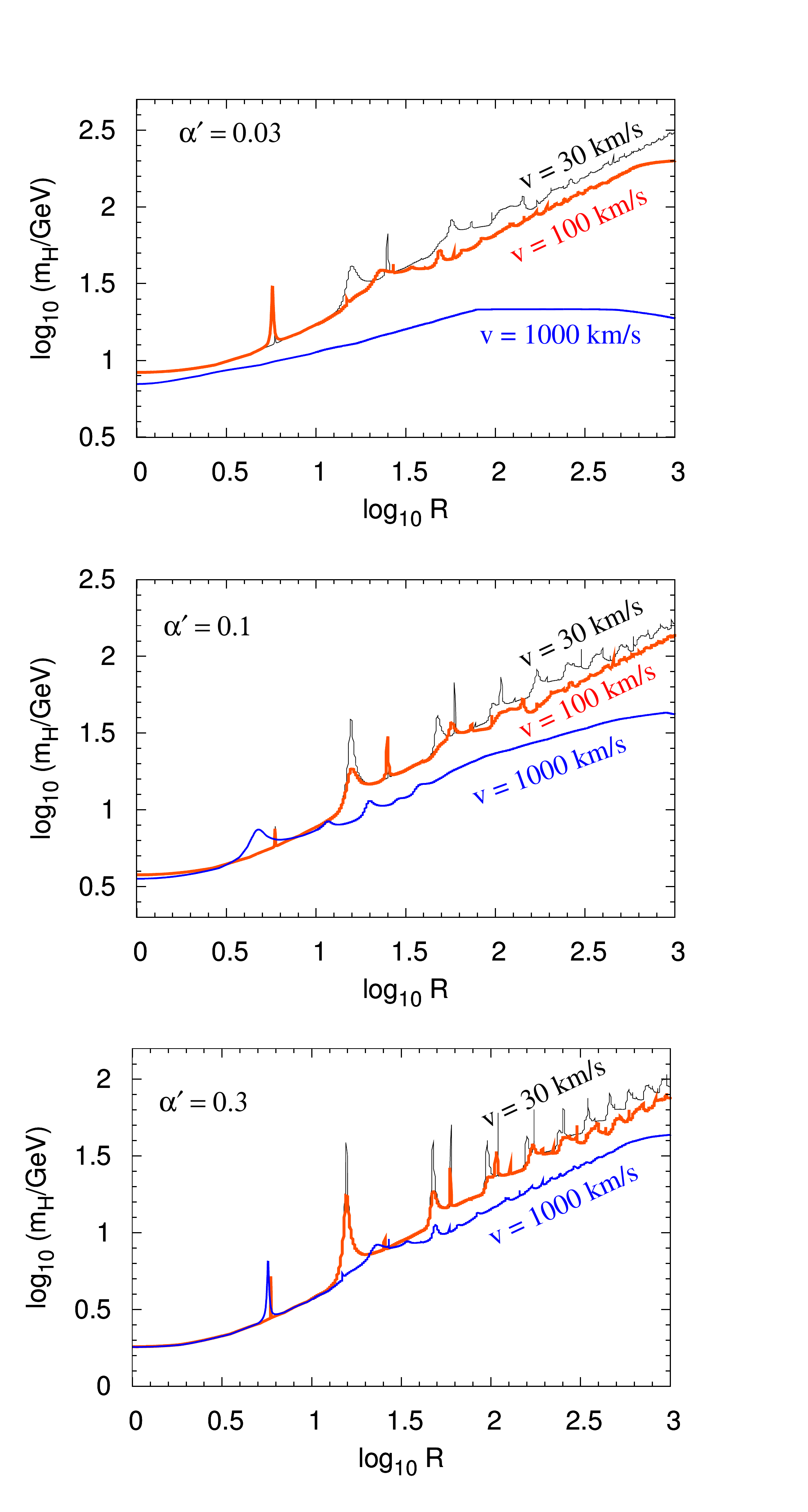}
$\!\!\!\!\!\!\!\!\!\!\!\!\!\!\!\!\!\!\!\!\!\!\!\!\!$
\includegraphics[width=1.1\columnwidth]{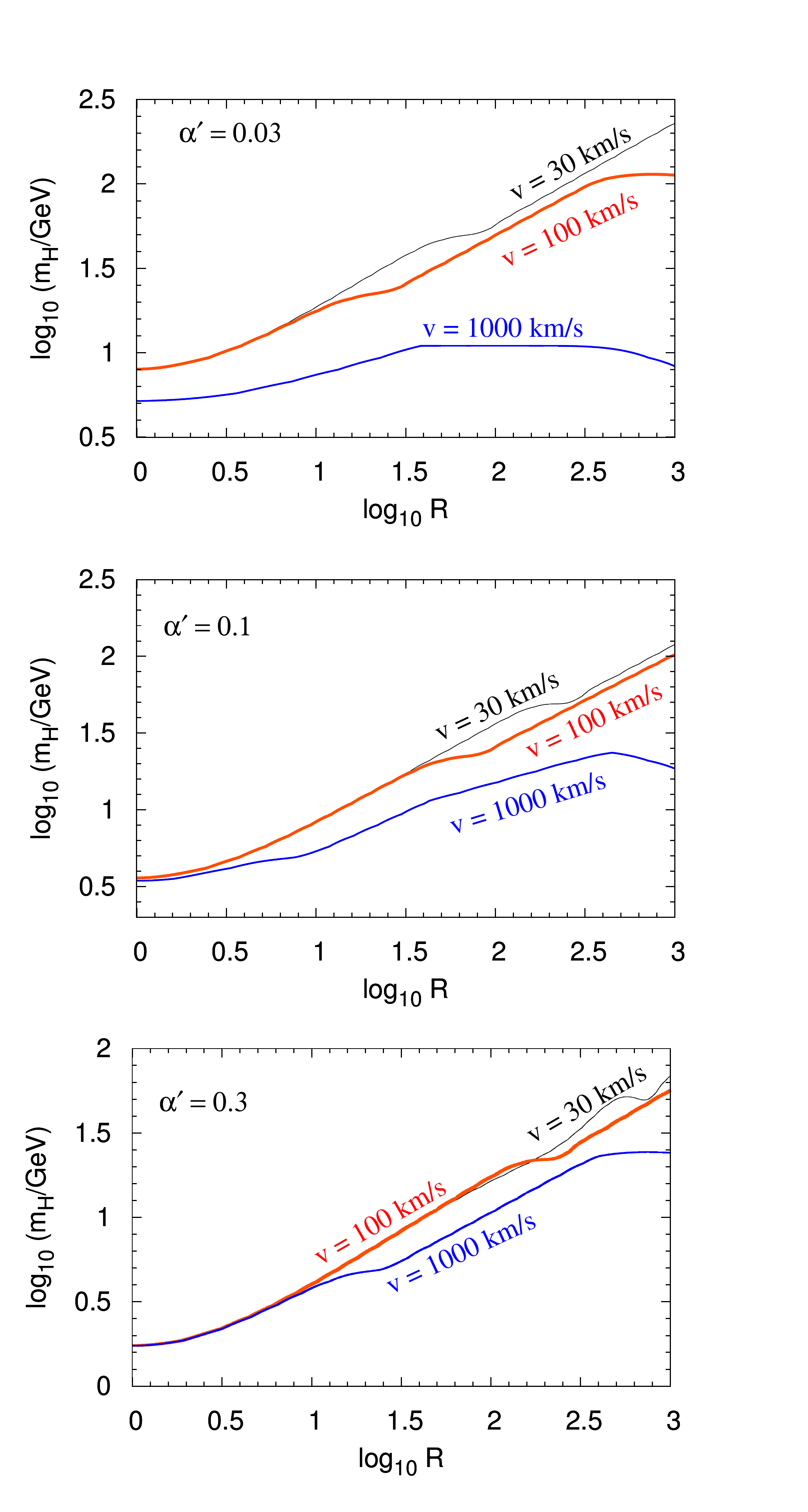}}
\vskip-0.5cm
\caption{Left: contours of constant $\sigma/m_\dH=0.6\,$cm$^2$/g 
in the plane of $m_\dH$ and $R=m_\pd/m_\de$ (using the atom-atom
momentum transfer cross section) at center of
mass energies $E = m_\dH v^2$, for $v=30$, $100$ and $1000$ km/s.
Top to bottom plots are for $\alpha'=0.03$, 0.1 and 0.3 respectively.
Right: analogous contours for molecular $\dH_2$ scattering,
with   $\sigma/(2m_\dH)=0.6\,$cm$^2$/g  and $E = (2m_\dH) v^2$, but
still using $m_\dH$ for the vertical axis.
}
\label{sidm-panel}
\end{figure*}

\subsection{Dark atoms}

For simplicity we will initially assume that dark atoms do not form a
significant population of molecules, but we will come back to this
question below.  We thus consider the elastic cross section for dark
atom scattering, which we have studied in detail in a previous paper
\cite{Cline:2013pca}.  In that work we computed both the elastic and momentum
transfer cross sections as a function of energy, over a large range 
of $R\sim 1-3000$, noting that the dependence upon $R$ can be very 
strong due to divergences of the scattering length in the channel
where the electrons are in the spin-singlet state.
The origin of these divergences can be understood from the 
form of the Schr\"odinger equation when rewritten using
atomic units of distance ($a_0'=1$) and energy ($\epsilon_0 \equiv 
\alpha'^2\mu_\dH=1$):
\be
	\left(\partial_r^2 -{\ell(\ell+1)\over r^2}
	+ f(R)\,(E-V_{s,t})\right) u_\ell^{s,t}(r) = 0
\label{Schr}
\ee
Here $u = r\psi$ and the subscripts $s,t$ label the spin-singlet and
triplet contributions to the scattering.
The singlet potential $V_s$
is much deeper than the triplet one $V_t$, and  it
rapidly acquires more bound states as $R$ is increased since the
potential is multiplied by $f(R)\sim R$.  Each time a bound state energy
approaches zero, the scattering length diverges.   From fig.\ 2
of ref.\ \cite{Cline:2013pca}, it can be seen that at low velocities,
the singlet channel typically dominates the scattering, except at
values of $R$ (such as in the real world) where the singlet scattering
length happens to be close to zero.

Using the methods described in ref.\ \cite{Cline:2013pca}, we have identified
the regions of atomic dark matter parameter space for which the 
momentum transfer cross section has the fiducial value $\sigma/m_\dH = 
0.6\,$cm$^2$/g.  The results are displayed as a function of $m_\dH$
and $R$ for several values of $\alpha'$ ($0.03,\,0.1,\,0.3$) in the left-hand
plots of fig.\ 
\ref{sidm-panel}.  We do not consider smaller $\alpha'$
since, as noted in \cite{Cline:2013pca}, then the ionization fraction starts to
become large in parts of the parameter space and the atomic
description is no longer appropriate.  We display contours of constant
$\sigma/m_\dH$ for several DM velocities, $v=30,\,100,\,1000$ km/s,
appropriate for dwarfs, Milky-Way-like galaxies and galactic
clusters, respectively.  Because the cross sections can
have significant velocity dependence, these curves do not generally 
coincide, although there are ranges of parameters where they do so,
namely for $R$ not too large and $\alpha'$ not too small. 

To better understand the results of fig.\ \ref{sidm-panel}, we recall
from ref.\ \cite{Cline:2013pca} that a typical scattering cross section for
dark atoms is of order $100\, a_0'^2$; therefore $\sigma/m_\dH \sim 
100\, \alpha'^{-2} f^2(R)\, m_\dH^{-3}$, implying that
\be
	{m_\dH\over{\rm\ GeV}} \cong \left({R\over
5.3\,\alpha'}\right)^{2/3}
\label{sidm_const}
\ee
for $R\gg 1$.  At fixed values of $R$ and
$\alpha'$, the higher curves in fig.\ \ref{sidm-panel}  require larger
$m_\dH$ to have the same cross section; therefore if $m_\dH$ was also
held fixed the lower curves would represent smaller values of
$\sigma$. DM distributions at the largest scales---clusters of
galaxies---which have the highest velocity dispersion, thus have the
smallest $\sigma$, except in the regions of large $\alpha'$ and
small $R$ where the curves overlap.  Where the curves do not overlap,
the cross section is generally largest for the smallest velocities,
but there are exceptions corresponding to resonances, which give rise
to the spiky structure as a function of $R$.  In particular, we find
that the bumps at $R\cong 5.6$ are due to a $p$-wave resonance in
the singlet channel (and thus do not appear in the scattering
length), similar examples of which are prominent in fig.\ 4 of
\cite{Cline:2013pca}.

Mirror symmetry, in which the dark sector is an exact copy of the
standard model, provides an explicit realization of atomic dark
matter, but one that is not compatible with the SIDM constraint
(\ref{sidm_const}), which requires that $m_\dH = 1.3$ TeV for the
SM values $R=1836$, $\alpha'=1/137$.   Extra freedom is possible in
the version of the model in which mirror symmetry is spontaneously
broken.  In this case the values $m_\dH \cong 5$ GeV and $m_\de \cong
50$ MeV have been promoted in a scenario where the visible and dark
baryon asymmetries are linked \cite{Berezhiani:2008zza}.  However
(\ref{sidm_const}) then requires $\alpha'\sim 1$, far from the 
value of $1/137$ that is still predicted despite mirror symmetry
breaking.

\begin{figure*}[t]
\centerline{\includegraphics[width=0.95\columnwidth]{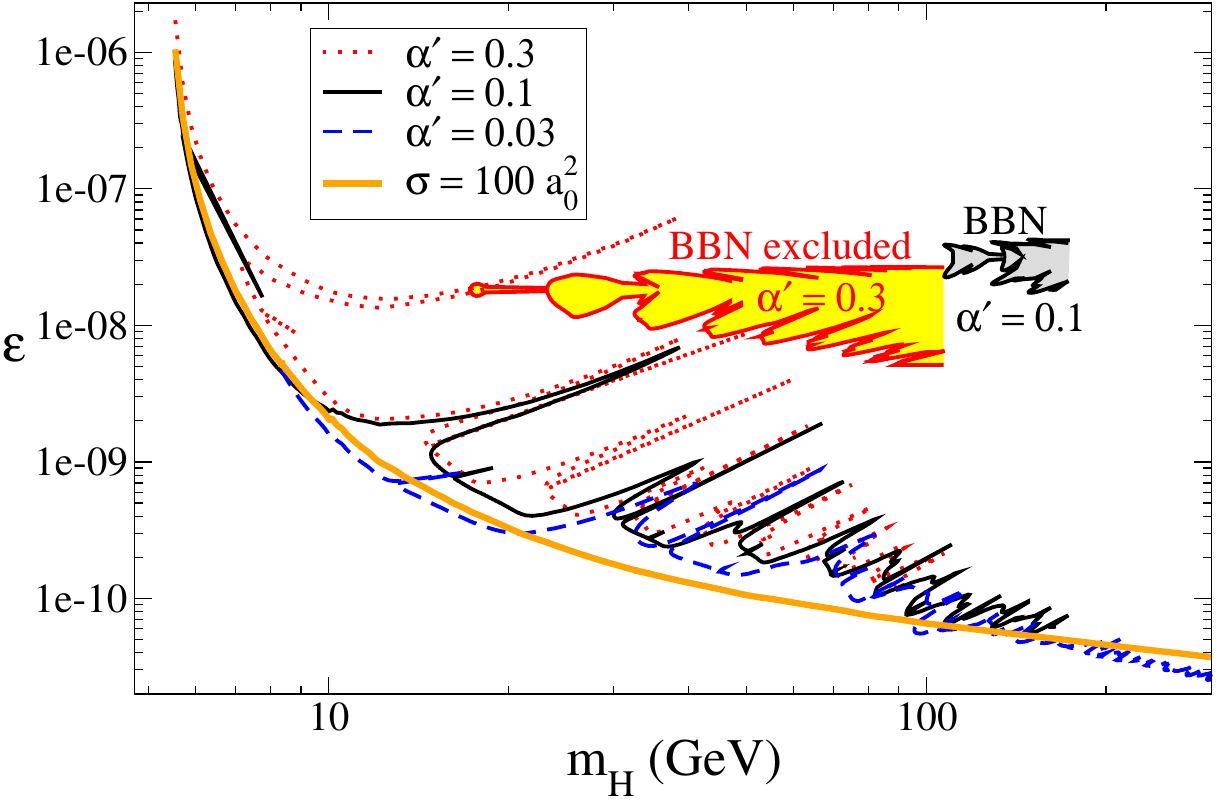}
\includegraphics[width=0.98\columnwidth]{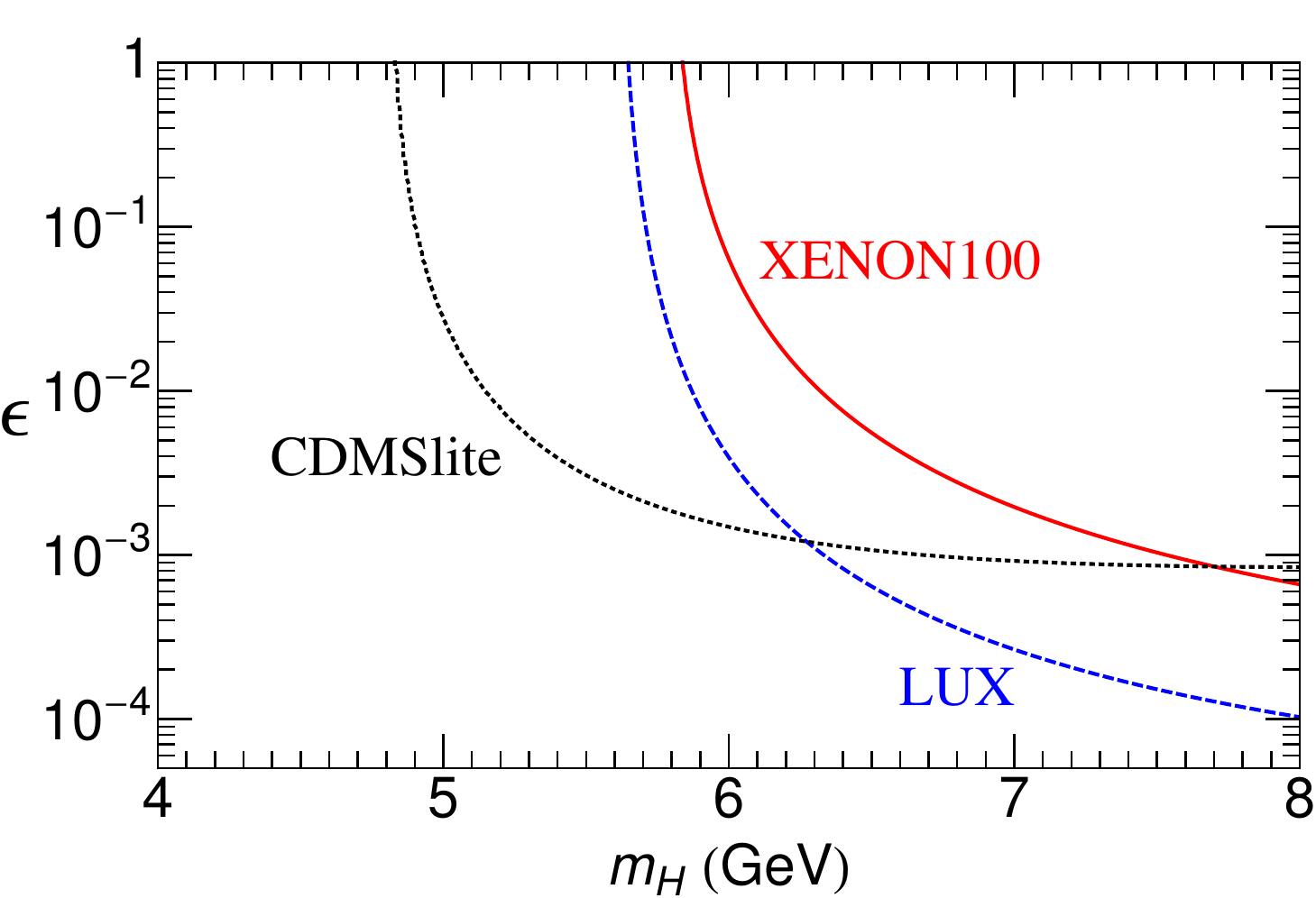}}
\caption{Left: upper limits on kinetic mixing in millicharged DM model from
 LUX \cite{Akerib:2013tjd} bounds, assuming the relation between $m_\dH$
and $R$ for atomic DM with $v=30$ km/s in fig.\ \ref{sidm-panel} is satisfied,
for each value of $\alpha' = 0.03,\,0.1,\,0.3$.  Also shown is the 
limit obtained from approximating the self-interaction cross section as
$\sigma = 100\, a_0^2$.  The shaded regions for indicated values
of $\alpha'$
are ruled out by big bang nucleosynthesis constraints, as described
in section \ref{bbnsect}.
 Right: constraints on $\epsilon$ for the special case 
$R=1$, in which the interaction of atomic DM with visible matter is through
an inelastic magnetic moment transition.  The DM mass splitting is determined
by the relationship (\ref{mdHrel}) that produces the target value of the
self-interaction cross section.
}
\label{eps-lim}
\end{figure*}

\subsection{Molecular dark matter}

In the interstellar medium, hydrogen gas exists not only in atomic form, but
also in H$_2$ molecules, whose abundance is significant especially in cold or
dusty regions where ionizing radiation is less present.  Although H$_2$ is
subject to destruction by ionizing radiation due to its relatively  weak
binding energy of 4.5 eV, it nevertheless requires ionized constituents such as
$p$ or $\HH^-$ for its formation, since the processes involving charged
particles, such as  $p + \HH\to \HH_2^+$ followed by $\HH_2^+ +\HH \to \HH_2
+p$,  are much more efficient for producing $\HH_2$ than the direct (but much
slower) process $H+H\to H_2+\gamma$.  Therefore the relative abundance of
molecules and atoms is not simple to predict.   Nevertheless (as we argued in
\cite{Cline:2013pca}) it seems plausible that molecules could be prevalent in a dark
sector where there are no stars, hence no ionizing radiation, since there is
still a small ionized fraction of the dark atomic constituents, of order
$f_i\sim 10^{-10}\alpha'^{-4} R^{-1} (m_\dH/{\rm GeV})^2$, that can give rise
to the catalyzed production of $\dH_2$.  A quantitative prediction of the 
$\dH_2$ abundance is beyond the scope of this paper.  Instead we consider the
prospects for dark molecules to have the desired self-interaction cross
section, assuming they constitute the dominant dark matter component.

The scattering cross sections of dark molecules were computed in  ref.\
\cite{Cline:2013pca}.  Using the methodology described there, we determined the
analogous constraints, from imposing that  $\sigma/m = 0.6\,$cm$^2/$g,  to
those of dark atoms and display them in the right-hand plots of fig.\  
\ref{sidm-panel}.
The general behavior of the curves is similar to their atomic counterparts
in fig.\ \ref{sidm-panel}, but the molecular ones are smoother as a function of
$R$, due to the shallow potential for molecule-molecule scattering, which does
not develop any bound state until $R\sim 700$.  Moreover, as also shown in
ref.\ \cite{Cline:2013pca}, the scattering length
for molecules has a zero near $R=280$, which gives rise to the pronounced dip
in the cross section at $v=10\,$km/s and $\alpha=0.1$.  This is a qualitative
difference with respect to the atomic case, where $\sigma$ at $v=10\,$km/s
is almost always larger than at $v=100\,$km/s.

\subsection{Direct detection constraints}
The model as presented so far does not give rise to any signal in
DM detectors, but by the simple addition of a kinetic mixing term
$\frac12\epsilon F^{\mu\nu}F'_{\mu\nu}$
between the photon and the dark U(1) gauge boson, it does so, as
was pointed out in ref.\ \cite{Cline:2012is}.  In that case the
dark electron $\de$ and proton $\pd$ acquire millicharges $\mp\epsilon e$ under
U(1)$_{\rm em}$, and so can scatter on protons by exchange of a
photon.  Even though the dark atom is electrically neutral, as long
as $R\gg 1$ the charge cloud of $\de$ does not overlap strongly with
that of $\pd$ and so $\dH$ will scatter on protons
electromagnetically, just like a normal H atom except for the reduced
charge of the dark constituents.  In the case of $R=1$, there is
strong cancellation between the two charge clouds and the Coulomb
scattering amplitude vanishes in the first Born approximation.
We will consider this special case separately.  

For $R\gg 1$, ref.\ \cite{Cline:2012is} showed that the cross section
for atomic DM scattering on protons is $\sigma_p = 4\pi
(\alpha\epsilon\mu_{p\dH})^2 a_0'^4$ where $\mu_{p\dH}$ is the reduced
mass of the $p$-$\dH$ system.  In the present context, we can fix the
value of $a_0$ for a given $m_\dH$ by assuming that the relationship
between $m_\dH$ and $R$ shown in fig.\ \ref{sidm-panel} is satisfied,
for given choices of $\alpha'$ and DM velocity.  Then $a_0 = f(R)/(\alpha'
m_\dH)$.  We choose $v=10$ km/s
since this tends to give the largest self-interaction cross section.
The maximum value of $\epsilon$ allowed by LUX \cite{Akerib:2013tjd}
can be found by setting the predicted $\sigma_p$ equal to the LUX
limit, relaxed by the factor $(A/Z)^2 = (131.3/54)^2$ due to the coupling
only to protons.  
In fact, the relation between $m_\dH$ and $R$ can be double-valued due to the
resonant peaks in $\sigma$, so we scan in $R$ to produce parametrized limit
curves in the plane of $\epsilon$ and $m_\dH$.  These are shown in fig.\ 
\ref{eps-lim}(left).  Except for the positions of the resonances, the $\alpha'$
dependence in these curves is weak.  For comparison we show the result of 
the rough approximation for the self-interaction cross section of 
$\sigma = 100\,a_0^2$, which results in the relation $a_0 = 5.3\, m_\dH^{1/2}\,
{\rm GeV}^{-3/2}$, which works well at low $m_\dH$ (apart from resonances), but gives somewhat too low
a prediction of $\sigma$ at higher $m_\dH$.

In sect.\ \ref{bbnsect} we will show that a small interval of
$\epsilon\sqrt{\alpha'}$  is excluded  over some range of dark
electron masses, (depending upon assumptions about initial conditions 
after reheating) to avoid overpopulating the dark photons during big
bang nucleosynthesis (BBN).  The excluded regions (shaded) 
in the $m_\dH$-$\epsilon$ plane are  shown
in  fig.\ \ref{eps-lim}(left), assuming that $m_\de$ is determined
by $m_\dH$ and $\alpha'$ so as to give the desired self-interaction
cross section.  For large values of $\alpha'\gtrsim
0.3$, this intersects part of the parameter space of interest
for direct detection, but for smaller $\alpha'$ there is no overlap
between the BBN-excluded regions and those which can be probed by
direct searches.

In the case of $R=1$, the leading interactions of dark atoms 
are, for sufficiently small $\alpha'$, the magnetic inelastic scatterings 
that change the total spin
of the atom.  These were studied in detail in ref.\ 
\cite{Cline:2012ei}, with attention to the region $m_\dH \sim 9-12$
GeV as suggested by excess events from the CoGeNT experiment
\cite{Aalseth:2010vx}.  For strongly interacting atomic DM satisfying
$\sigma/m_\dH = 0.6\,$cm$^2/$g, we find that $m_\dH$ and $\alpha'$ 
are related by 
\be
	m_\dH = 0.8\,\alpha'^{-2/3}\,{\rm GeV}
\label{mdHrel}
\ee 
so that $m_\dH$
ranges between 1.8 and 8.3 GeV for $\alpha' = 0.3-0.03$.  The rate of
nuclear scatterings is sensitive to the mass difference $\delta m_\dH$
between the lowest 
DM state and the hyperfine excitation where the relative spins of the proton
and electron are reversed, $\delta m_\dH = \alpha'^4 m_\dH/6$, from which we
can eliminate $\alpha'$ in favor of $m_\dH$ using (\ref{mdHrel}), for
the case of self-interacting DM. 

\begin{figure*}[t]
\centerline{\includegraphics[width=0.98\columnwidth]{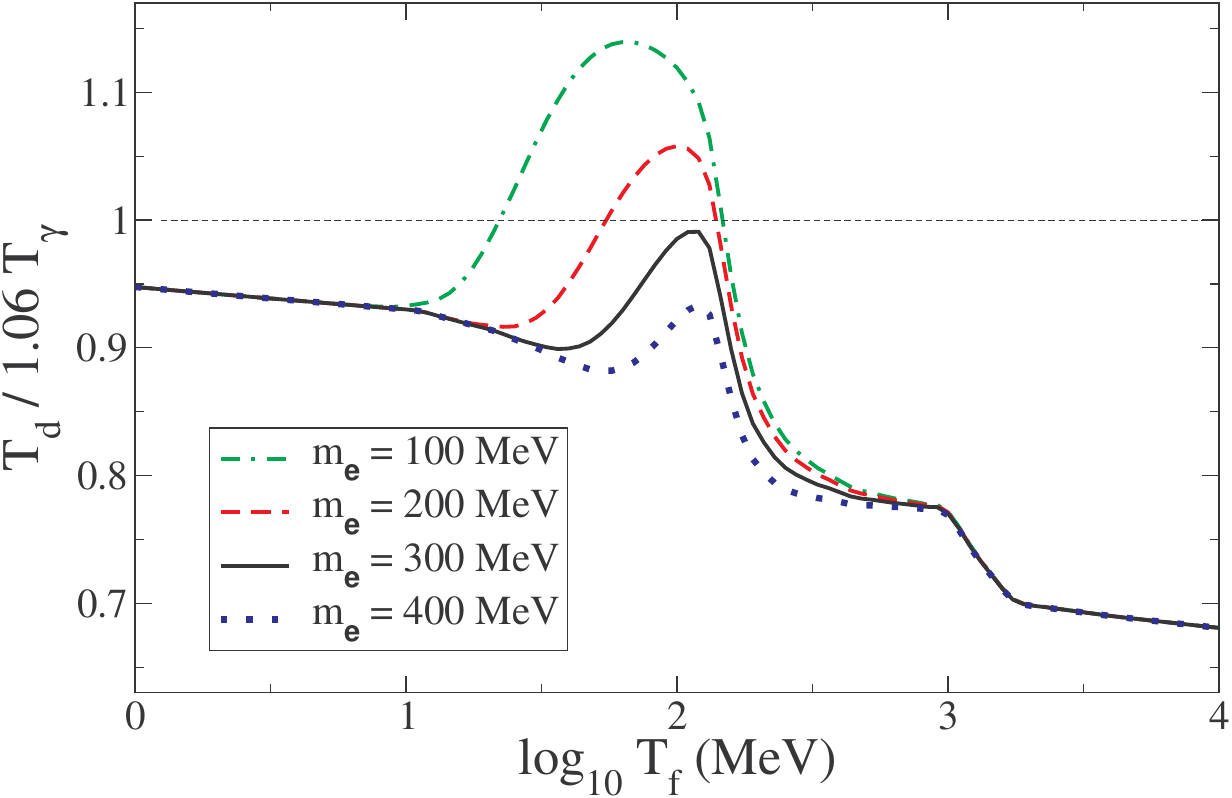}
\includegraphics[width=\columnwidth]{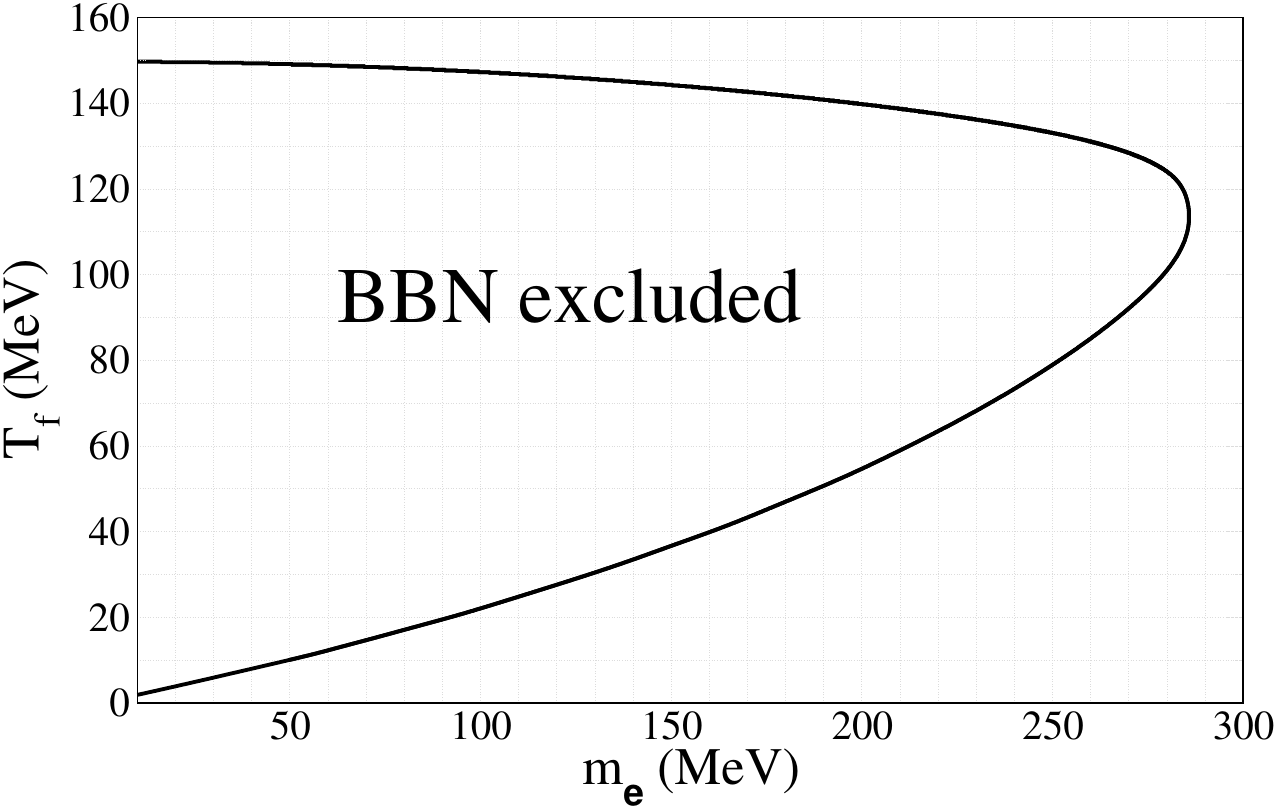}}
\caption{Left: Dark to visible temperature ratio divided by the 
limiting value 1.06, as determined by eq.\ \ref{Tdeq}, as a function
of the freeze-out temperature $T_f$ for the process 
$\gamma' \de \leftrightarrow \gamma \de$.  Relation is shown for
several values of $m_\de$.
 Right: the resulting excluded region (shaded) in the  $m_\de$-$T_f$ 
plane.
}
\label{bbn}
\end{figure*}

In  fig.\ \ref{eps-lim}(right) we show the constraints on the kinetic mixing
for the $R=1$ model from CDMSLite \cite{Agnese:2013jaa}, XENON100
\cite{Aprile:2012nq} and LUX \cite{Akerib:2013tjd}, for models that satisfy
(\ref{mdHrel}) and therefore have the required self-interaction cross 
section.\footnote{The XENON100 limit is computed as described in appendix D of 
\cite{Cline:2012ei}.  For the LUX limit, the number of events is
 computed using the acceptance function given in fig.\ 9 of
\cite{Akerib:2013tjd} (black `$+$' symbols). 
The 90\% upper limit on the events due to DM, $N<2.4$ events, is
used for low mass  DM. Following LUX, the events below 3 keVnr are not 
included.  For CDMSlite, which has not made its data publicly available, we
randomly distributed
events within histogram bins in the energy range 841 eVnr (nuclear recoil threshold) to
4 keVnr (avoiding the activation line near 5.3 keVnr). 
For low mass DM the exclusion limit is essentially controlled by the
low recoil energy spectrum, so the 90\% limits are computed using the
$p_{\rm Max}$ method \cite{Yellin:2002xd}
in the nuclear recoil energy range [0.841, 4] keVnr. Although for elastic DM,
the CDMSlite limits extends to much lower DM mass range, the limits for 
millicharged atomic dark matter
terminate near $m_H = 5$ GeV due to inelasticity.}  
The mass splitting $\delta m_\dH$ ranges from 14 to 1 keV for $m_\dH = (5-8)$
GeV.  The constraints on
$\epsilon$ are much weaker than in the case of $R \gg 1$ due to the 
inelastic magnetic
dipole versus elastic Coulomb nature of the interaction.  We do not show the
CoGeNT-allowed region on fig.\ \ref{eps-lim} because we find that the
assumption (\ref{mdHrel}) needed to have strongly-interacting DM is contrary to
getting a good fit to the CoGeNT excess events.  If we nevertheless 
impose (\ref{mdHrel}), the best fit region is at higher masses than allowed for
strongly interacting atomic dark matter with small ionization fraction, with the edge of the 99.7\% C.L.\ region just reaching the right-hand
side of the plot.  There is essentially no overlap between the CoGeNT events
and the SIDM parameter space within this model.

\begin{figure*}[t]
\centerline{\includegraphics[width=0.98\columnwidth]{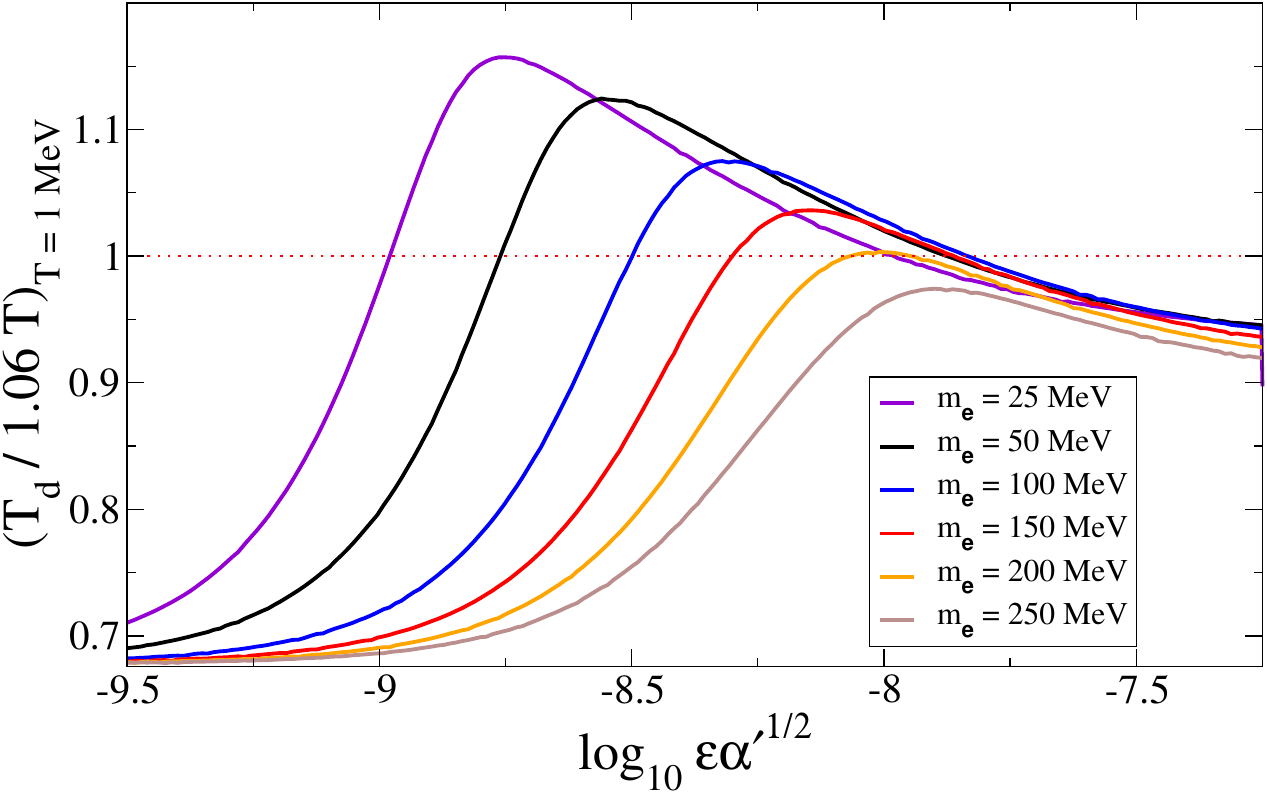}
\includegraphics[width=0.98\columnwidth]{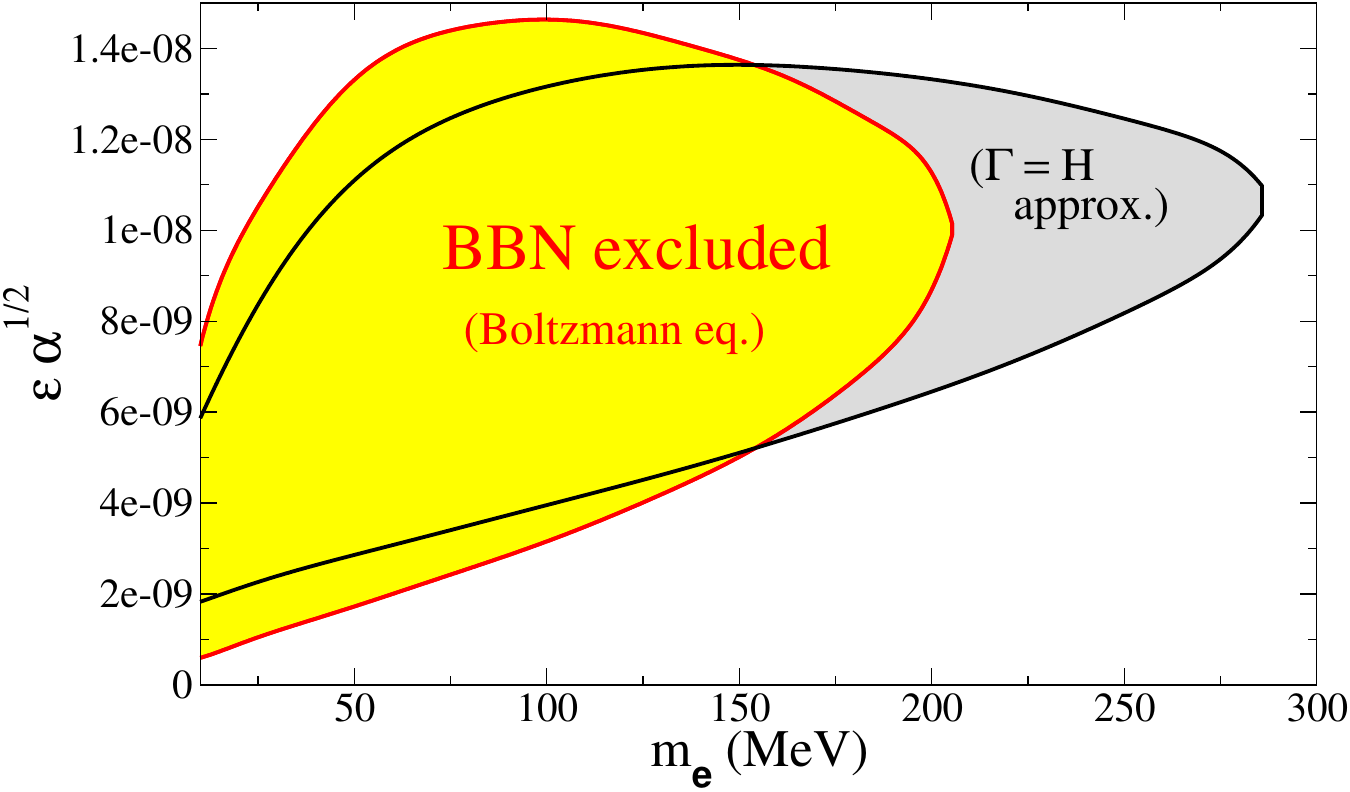}}
\caption{Left: dark to visible photon temperature ratio $T_d/T$ at 
$T=1$ MeV versus $\epsilon\sqrt{\alpha'}$, as determined by 
Boltzmann equations (\ref{boltz}), for
several values of the dark electron mass (curves are in order of
increasing $m_\de$ from top to bottom as in legend).  The minimum
value of $\epsilon\sqrt{\alpha'}$ shown for a given $m_\de$ corresponds
to the critical value $(\epsilon\sqrt{\alpha'})_c$ defined in the text.
Right: BBN excluded regions of $\epsilon\sqrt{\alpha'}$  versus $m_\de$
for both the simple estimate based on eq.\ (\ref{Tdeq}) and the 
Boltzmann code.
}
\label{bbn2}
\end{figure*}

\subsection{Big bang nucleosynthesis constraints}
\label{bbnsect}

The atomic DM model can potentially provide an excess of radiation
during the epoch of big bang nucleosynthesis (BBN).  This provides
stringent constraints on mirror dark matter 
\cite{Carlson:1987si}-\cite{Berezhiani:2008gi} since both the dark
photons and dark electrons/neutrinos can contribute to the excess.  In our case,
$m_\de$ is always greater than $100$ MeV so that only the dark photons
can contribute (and we do not consider dark neutrinos).   Taking the
95\% c.l.\ limit on the effective number of extra neutrino species
$\delta N_\nu < 1.44$ \cite{Cyburt:2004yc}, 
the dark
photon temperature $T_d$ at the time of BBN is constrained
to be $T_d/T =  (\frac78\delta N_\nu)^{1/4}< 1.06$ 
relative to that of the 
visible photons.
If there is no interaction between the dark and
visible sectors, then the excess in dark radiation depends upon
initial conditions, and can be compatible with the constraints if
reheating into the dark sector after inflation is less efficient than
into the visible one. 

An interesting scenario is that where gauge kinetic mixing between the
dark and visible photons provides an interaction between the two sectors,
causing the dark proton and electron to have millicharges $\epsilon e$
under electromagnetism.  This can lead to equilibration between the dark
and visible photons through scattering on  dark electrons, $\gamma' \de
\leftrightarrow \gamma \de$.    At low temperatures, the interaction rate
is governed by the
Thomson cross section $\sigma = (8\pi/3)
\alpha\alpha'(\epsilon/m_\de)^2$, but for $T> m_\de$ one must perform
the thermal average of the Compton cross section as described in 
appendix \ref{bbnapp}.  
This interaction  can come into equilibrium
at temperatures above $\sim\!\!m_\de$ if $\sqrt{\alpha'}\epsilon$ does
not fall below a critical value that we find to be given by
$(\sqrt{\alpha'}\epsilon)_c\sim 10^{-10.6} (m_\de/{\rm MeV})^{0.57}$,
assuming that $T_d=T$.  It goes back out of equilibrium at lower
temperatures as the dark electrons disappear from the bath.

Although it is beyond the scope of this paper to make a detailed 
study of the thermal history of the dark sector, it seems reasonable
to suppose that $T_d = T$ at some early time, if $\epsilon$ is 
not too small.  For example, even if the dark sector was initially
much colder than the visible one, say at the moment of reheating,
the interaction $\gamma\gamma\to\de^+\de^-$ comes into equilibrium 
by $T\sim m_\de$ if $\alpha\epsilon \gtrsim (m_\de/M_p)$.  We will be
interested in $m_\de\sim 100$ MeV, for which this implies
$\epsilon\gtrsim 10^{-8}$, compatible with the magnitude for which we will
find BBN constraints.  
For the following discussion, we assume that $T_d=T$ at
temperatures of a few GeV as an initial condition.

Under these assumptions, 
If $\gamma' \de \leftrightarrow \gamma \de$ goes back out of
equilibrium at the wrong time, there is a risk that the entropy dumped
into the dark photon bath from $\de\de\to\gamma'\gamma'$ will conflict
with BBN constraints on the total radiation density.  There is a
competition between the heating of the $\gamma'$ bath versus 
the heating of the visible photons during the QCD phase transition. If
$\epsilon$ is very small, the freeze-out of $\gamma' \de \leftrightarrow \gamma \de$
occurs at such a high temperature $T_f$  that the dark sector
decouples before the QCD transition heats the visible sector, and so
the dark photon temperature $T_d$ is suppressed relative to the
visible $T$.  If $\epsilon$ is
very large, the two baths remain coupled down to low temperatures, 
so that dark electron annihilations heat all lighter degrees of 
freedom equally.  In this case $T_d$ also does not exceed the 
visible $T$.  Therefore we expect only a limited
range of $\epsilon$ to be excluded by BBN.

We will determine the BBN constraint on $\epsilon$ in two different 
ways, one simpler and
the other more quantitative.  In the first estimate, 
using entropy conservation, the ratio of the dark to visible photon
temperatures at the time of $\gamma' \de \leftrightarrow \gamma \de$
freeze-out is 
\be
	{T_d\over T} = \left[\left({2 + g_\de(x_f)\over
2}\right) 
	\left({10.75\over g_*(T_f)}\right)\right]^{1/3} < 1.06
\label{Tdeq}
\ee
where $x_f = m_\de/T_f$ and $g_\de(x_f)\le 3.5$ is the 
effective number of dark 
electron degrees of 
freedom in the plasma at the freezeout temperature $T_f$, 
\be
	g_\de(x) = {45\over (\pi T)^4}\int_0^\infty dp\, {p^2(E+p^2/3E)
	\over e^{E/T}+1}
\ee
while $g_*$
is the effective number of entropy degrees of freedom in the 
standard model at $T_f$.  We use the results of 
recent lattice studies of the QCD phase transition
\cite{Borsanyi:2010cj} for $g_*(T)$.  In fig.\ \ref{bbn} we show
how eq.\ (\ref{Tdeq}) depends upon $T_f$ for a range of $m_\de$,
and the resulting excluded values of $T_f$ versus $m_\de$.
Notably, for $m_\de\gtrsim 285$ MeV, there is no constraint in 
this approximation, since
the dark electrons annihilate early enough for their effects to 
be counteracted by the QCD phase transition.

By equating the $\gamma' \de \leftrightarrow \gamma \de$ scattering
rate $\Gamma = n_\de\langle\sigma v\rangle$ to the Hubble rate, we
find the freeze-out temperature $T_f$ as a function of $m_\de$ and 
$\alpha'\epsilon^2$.  Combining this with the excluded values of $T_f$
versus $m_\de$, we obtain a corresponding constraint on 
$\epsilon\sqrt{\alpha'}$ to avoid the nucleosynthesis bound.  The
result is shown in fig.\ \ref{bbn2}.

For a more quantitative determination of the BBN constraint, we solve
the coupled Boltzmann equations for the energy densities of the dark
and visible photons,
\bea
	{d\rho_{\gamma'}\over dt} &=& -4H\rho_{\gamma'} + q_{\rm ann} + 
	q_{\rm scatt}\nonumber\\
		{d\rho_\gamma\over dt} &=& -4H\rho_{\gamma\phantom{'}} + q_{\rm\sss SM}  
	\,-q_{\rm scatt}
\label{boltz}
\eea
where the source terms $q_i$ are due to $\de\de\to\gamma'\gamma'$
annihilation, $\gamma' \de \leftrightarrow \gamma \de$ scattering,
and $\bar f f\to \gamma\gamma$ annihilation of standard model
particles $f$, respectively.  Details are given in appendix
\ref{bbnapp}.  We obtain $T_d/T = 
(\rho_{\gamma'}/\rho_{\gamma})^{1/4}$ as a function of $T$ in this
way, and evaluate it at $T=1$ MeV appropriate for BBN, demanding that
$T_d/T$ not exceed 1.06.  The results are qualitatively similar to those
of the simpler estimate, but slightly  more constraining in $\epsilon$ while
less so in $m_\de$.  Fig.\ \ref{bbn2} shows that constraints exist for 
$m_\de$ up to 200 MeV, in contrast to the value 285 obtained
previously. This
limit is used, along with the relation between $m_\de$
and $m_\dH$ from fixing the self-interaction cross section,
 to construct the BBN-excluded regions shown in 
fig.\ \ref{eps-lim}(left).  We find that in general, direct detection
provides a stronger constraint than BBN for this model.

\subsection{Structure formation and CMB constraints}

Refs.\ \cite{Feng:2009mn},\cite{CyrRacine:2012fz}  studied various cosmological
constraints on atomic dark matter.  There it is shown that the power
spectrum of matter fluctuations is suppressed at small scales because
of the analog of baryon acoustic oscillations, unless recombination in
the dark sector occurs sufficiently early.  This puts an
upper 
bound on the interaction strength $\alpha'$, but one that is easily
satisfied by our models of interest for SIDM. 

The main observational constraint is that the matter power spectrum
should not differ from the $\Lambda$CDM prediction at scales $k <
2h$Mpc$^{-1}$, based on Ly-$\alpha$ measurements.  This must be
compared to the scale at which dark atom acoustic oscillations start
to occur, given by the dark sector sound horizon $r_d$ at the time of
its kinetic coupling: $r_d = \int_0^{a_{\rm dec}} c_d/(Ha^2) \cong
(T_0/T_{\rm dec})/(\sqrt{3} H_0)$, where $c_d$ is the dark sector
sound speed, which we roughly approximate as $1/\sqrt{3}$ until the
time of decoupling, and zero afterwards. The kinetic decoupling
temperature $T_d$ is determined in terms of the atomic DM model
parameters in \cite{CyrRacine:2012fz}, resulting in the lower bound
\be
	{T_d\over E_b} > {6\times
10^{-13}\over\alpha'^6\,\zeta^4}\left({E_b\over {\rm\ keV}}\right)\left({
	m_\dH\over{\rm\ GeV}}\right)
\ee
where $E_b$ is the binding energy and $\zeta$ is the ratio of
temperatures in the dark and visible sectors.  We thus find that a sufficient
condition to satisfy the Lyman-$\alpha$ bound is 
\be
	\alpha' < {34\over\zeta^2}\left({m_\dH\,\mu_\dH^2\over{\rm\ GeV}^3}
	\right)^{1/2}
\ee
where $\mu_\dH = m_\dH/f(R)$ is the dark electron reduced mass.
This is satisfied for all models obeying the SIDM constraint in 
fig.\ \ref{sidm-panel}.

Cosmic microwave background constraints on atomic dark matter models
were summarized in ref.\ \cite{Cline:2012is}.  The main requirement is
that the dark atoms be out of kinetic equilibrium with the
baryon-photon plasma before recombination. This would lead to
stringent constraints on the kinetic mixing parameter $\epsilon$ from
Rutherford scattering if the dark atoms were ionized 
\cite{McDermott:2010pa}, but if the ionization fraction is negligible
(as we demand) then the relevant process is Rayleigh scattering 
of visible photons into dark photons, which
is much weaker.  (The scattering of visible photons into themselves is
weaker still, suppressed by a further factor of $\epsilon^2
\alpha/\alpha'$.)  
The cross section is given by
\be
	\sigma_R = \epsilon^2{\alpha\over\alpha'}\sigma_T (E_\gamma/E_b)^4
\ee
where $\sigma_T = 8\pi\alpha'^2/(3 m_\de^2)$ is the dark Thomson cross
section and $E_b$ is the binding energy.  The rate of photon-atom
interactions is $\Gamma \sim n_\gamma \sigma_R$ with $n_\gamma = 
0.24\, T^3$ at the recombination temperature $T_r = 0.26$ eV.  Demanding
that $\Gamma(T_r) < (3.8\times 10^5$y)$^{-1}$, and taking
$E_\gamma\cong 2.7\,T_r$, we find that the constraint is satisfied by
orders of magnitude even if $\epsilon\sim1$.  
The same conclusion is true for the new CMB bounds on photon-dark matter
scattering  found by ref.\ \cite{Wilkinson:2013kia}.  We find that
$\epsilon <  (\alpha'/0.01)^{7/2} (m_\de/{\rm\ MeV})^{7/2}$, but
$m_\de$ is $\gtrsim$ 0.1 GeV for the SIDM models shown in fig.\
\ref{sidm-panel}.  Such large values of $m_\de$ imply that dark photons
freeze out (via dark electron annihilation) at sufficiently high 
temperatures so that there is no danger of producing a too-large density
of dark radiation.

Recently ref.\ \cite{Cyr-Racine:2013fsa} refined the constraints 
from dark acoustic oscillations, which could reveal the effects
of dark atoms on large scale structure even in the absence of any
nongravitational interaction between the two sectors.  
There the constraint 
\be
    \Sigma_{\rm DAO} = 2\times 10^{-9}\alpha'^{-1} f(R)
	\left(m_\dH\over{1{\rm\ GeV}}\right)^{-7/6} <
10^{-4}
\label{DAO}
\ee
is derived (in which we have made explicit the dependence of the
binding energy on $\alpha'$ and the reduced mass $\mu_\dH$, and 
used (\ref{fRrel}).  Using (\ref{sidm_const}) to eliminate 
$\alpha'^{-1} f(R)$, we find that (\ref{DAO}) is satisfied as long
as $m_\dH < 10^{12}$ GeV, which has no effect on our preferred
parameter space.

\section{``Mesonic'' dark matter}
\label{sec:meson}

Composite dark matter could be analogous to hadronic states of
QCD if it is bound by a confining force associated with a 
nonabelian gauge symmetry.  The most natural choice would be the
baryonic states of such a theory, since it is the baryons of 
QCD that are stable in the visible sector.  However, this is not
the only possibility.  If there is no analog of weak or
electromagnetic interactions in the dark sector, then mesonic
bound states could be stable, and if lighter than the baryons,
could constitute the dark matter.  Glueballs of the dark sector might
alternatively be the dark matter, in the case where the constituent
particles are heavier than the confinement scale, or absent
altogether.  We consider the mesonic case in this section, and the baryonic/glueball cases
in successive ones.

\subsection{Elastic scattering cross section}

If there are ``quarks'' transforming in the fundamental representation
of a dark sector SU($N$)$_d$, they will form
mesonic $q\bar q$ bound states that could be stable or metastable
bosonic dark matter candidates.  Below the confinement scale, one
expects that the elastic scattering interaction for such mesons
will be strong, possibly fulfilling the criteria for SIDM.  The
$2\to 2$ low-energy elastic scattering amplitude can be estimated 
using chiral perturbation theory (for a review, see ref.\
\cite{Georgi:1985kw}), with the Lagrangian 
\be
	{F_\pi^2\over 4}\tr\left( \partial_\mu\Sigma^\dagger\,
	\partial^\mu\Sigma\right) + {\xi\over 4} F_\pi^3\,\tr  \left(M\Sigma + {\rm
	h.c.}\right)
\label{chiral_lag}
\ee
where $\Sigma = e^{2i{\Pi}/F_\pi}$,  $F_\pi$ is the analog of the pion
decay constant ($F_\pi=93$ MeV for QCD), $\xi$ is $O(1)$, and $M$ is the
dark quark mass matrix.  For simplicity we take $M=m_q{\mathbf 1}$, proportional to
the unit matrix.   If there are $N_f$ flavors of dark quarks, then
$\Pi$ is a  $N_f\times N_f$ matrix  given by $\Pi = \pi^a T^a$, where
$T^a$ are the generators of SU($N_f$), normalized such that $\tr T^a
T^b = \frac12\delta_{ab}$.  The pion mass is given by $m_\pi^2 = \xi F_\pi
m_q$.

{}From the $\pi\pi\to\pi\pi$ elastic scattering amplitude, we
obtain the cross section (see appendix \ref{chipt} for details)
\be
	\sigma = { m_\pi^2\over 32\pi\,F_\pi^4}\,C(N_f)
\ee
where $C(N)=(2N^4-25N^2+90-65/N^2)/(N^2-1)$.
Taking for example $m_\pi \cong 1.5 F_\pi$ as in QCD, 
this gives $\sigma \cong 0.05\, C(N_f)/m_\pi^2$.  To achieve the desired
SIDM cross section of $\sigma/m = 0.6\,$cm$^2$/g then requires
$m_\pi = (33,\,36,\,61,\,83,\,100)$ MeV for $N_f=(2,\dots,6)$, respectively.
For general choices of the ratio $x=m_\pi/F_\pi$, which depends
upon the dark quark mass as $m_q^{1/2}$, these values
should be rescaled by $(x/1.5)^{4/3}$.
Unlike an elementary boson of such a small mass,
there is no naturalness problem in principle because the mass scale
is not fundamental here, but is determined by the running of the dark
SU($N$) gauge coupling through the confinement scale $\Lambda_d\sim 4\pi
F_\pi$, and the quark mass $m_q$ whose smallness is protected by chiral
symmetry.

\subsection{Relic density constraints}

An example providing stable ``pionic''
dark matter was recently proposed \cite{Bhattacharya:2013kma}, in
which the presence of two quark flavors with isospin symmetry assures
the stability of the pions.  In that model, nonrenormalizable
interactions between the pions and the standard model were invoked 
so that $\pi\pi$ annihilations to SM states result in the observed
relic density.  These were assumed to be mediated by exchange of the Higgs or 
the linear combination of $\gamma$ and $Z$ corresponding to weak hypercharge,
through the interactions
$\lambda_h\,|H|^2\tr(\partial_\mu\Sigma^\dagger\partial^\mu\Sigma)$ and
$\lambda_v\,B^{\mu\nu}\tr(\Sigma^\dagger\partial_\mu\Sigma\partial^\nu\Sigma^\dagger)$
respectively.\footnote{The latter operator (and those in eqs.\ (14,16)) breaks chiral symmetry; it
implicitly contains a quark mass matrix insertion}\ \  It was shown that by fixing $\lambda_h$ or $\lambda_v$
to give the right relic density, and imposing constraints from the
invisible decays $H\to\pi\pi$ and $Z\to\pi\pi\pi$, the decay constant
$F_\pi$ should be greater than several times 10 GeV, which is much larger
than needed for the SIDM cross section as we estimated in the previous
section.

One is therefore led to question: do there exist any possible forms
of  interactions between the dark pions and the standard model that
would allow for standard thermal production through annihilation of
$\pi\pi$ into known particles, while respecting stability of the $\pi$
and not conflicting with particle physics constraints?  A few examples
suffice to show that any mediator interactions between $\pi$ and
standard model fermions $f$, expressed as higher
dimension operators such as
\be
	{ m_\pi^2 F^2\over M^4}\, \tr(\Sigma+\Sigma^\dagger)\left[
	\bar f h f,\ \bar f\slashed{\partial} f,\ \dots\right]
\ee
whose strength is consistent with the desired
relic density of dark pions,  are suppressed by a very low scale in
the denominator, $M\lesssim 1$ GeV.   (If there are additional small
couplings due to approximate symmetries then $M$ must be even
smaller.)  
Hence a thermal origin of pionic DM of
such low masses requires new physics below the GeV scale coupling the
dark sector to the standard model.  Such models have been explored 
in connection with SIDM in ref.\ \cite{Tulin:2013teo,Kaplinghat:2013kqa}.
Once such light mediators are admitted, the motivation to invoke
compositeness to explain the strong self-interactions might be
diminished, since light mediators are already sufficient for that
purpose.  However if the particles in the dark sector that interact
with the light mediators are composite, the situation is qualitatively
different from those that were previously considered.  We outline 
such a model for light pionic DM in the next subsection.

Another possibility that admits much weaker interactions between
$\pi$ and the visible sector is for $\pi$ to be metastable with
respect to the age of the universe.   For example,
$\pi$ could decay into light SM fermions analogously to the weak
interactions, by mixing with a superheavy $Z'$, giving a 
lifetime of order $m_{Z'}^4/m_\pi^5$.  For $m_\pi\sim 100$ MeV,
observations of the isotropic diffuse $\gamma$-ray background 
constrain the lifetime to be $\tau > 5\times 10^{24}$ s for
$\pi\to e^+e^-$ \cite{Bell:2010fk}.  Distortions of the cosmic 
microwave background give a stronger limit,
$\tau > 5\times 10^{25}$ s \cite{Diamanti:2013bia}, requiring $M_{Z'}\gtrsim
10^{11}$ GeV.  Even if $\pi$ decays only into neutrinos, the
limit on $M_{Z'}$ is relaxed by a factor of just 2.6. 

Given such feeble interactions, one could  try to use the
``freeze-in'' mechanism of ref.\ \cite{Hall:2009bx} as an alternative
means of thermally producing its relic abundance.   By this
mechanism, the relic abundance of $\pi$ is predicted to be of order
\be
	Y_\pi \sim {m_\pi^3\, M_p\over M_{Z'}^4}
\label{fimp}
\ee
while the required value for the relic abundance is given by 
$Y_\pi =4\times 10^{-9}(m_\pi/100{\rm\ MeV})^{-1}$.  Combining this
with eq.\ (\ref{fimp}) gives $M_{Z'} \cong 10^6$ GeV $\times(m_\pi/100{\rm\
MeV})$, in strong conflict with the diffuse $\gamma$-ray or
neutrino constraints.
In the following we construct a model with light mediators that
circumvents these constraints.

\subsection{Model of light pionic DM}

We can devise a model of light, strongly interacting pionic dark 
matter that has the desired thermal
relic abundance, if the dark sector contains a broken 
U(1)$'$ gauge symmetry with a sufficiently light $Z'$ gauge boson,
that mixes kinetically with the photon.  The model is similar to
that of ref.\ \cite{Bhattacharya:2013kma}, but instead of coupling
the pions to the $Z$ boson, we couple them to the $Z'$.  If all the
dark quarks have the same U(1)$'$ charge (as well as equal masses) then
the diagonal (vector) SU($N_f$) flavor symmetry remains unbroken.
The lowest dimension operator consistent with this symmetry, that couples
the pions to the $Z'$, is
\be
	{\lambda_0\, m_q\over F_\pi}\,Z'_{\mu\nu}\tr\left(\Sigma\,
	\partial^\mu\Sigma^\dagger \partial^\nu\Sigma\right) +{\rm h.c.}
\label{zprime}
\ee
This can lead to freeze-out of the pions through the coannihilation process
$\pi\pi\to \pi Z'$.  However we find that the matrix element is highly velocity
suppressed ($d$-wave, details in appendix \ref{pi-ann}).  The relic density is thus
more likely to be determined by the higher dimension operators
\be
{\lambda_1\over 4F_\pi^2}\, Z'_{\mu\nu}{Z'}^{\mu\nu}\,  
\tr(\partial_\alpha\Sigma^\dagger
\partial^\alpha\Sigma)
+{\lambda_2\over 4F_\pi^2}\, Z'_{\alpha\mu}{Z'}^{\nu\alpha}\,\tr(\partial^\mu\Sigma^\dagger \partial_\nu\Sigma)
\ee
that give rise to $\pi\pi\to Z'Z'$.  Ignoring for simplicity the interference
between these two operators, we find that the corresponding cross sections are 
given by 
\be	
	\sigma v = {m_\pi^6\over \pi F_\pi^8}\left\{8\lambda_1^2,\
\frac32\lambda_2^2\right\}
\label{sigma-ann}
\ee
as $v\to 0$.  Taking $F_\pi \cong 0.67 m_\pi\cong 24$ MeV (using the relation
between $F_\pi$ and $m_\pi$ for QCD and the value of $m_\pi$ needed for 
SIDM in the case $N_f=3$), and using the relic density cross section $\sigma v = 
4.5\times 10^{-26}$ cm$^3$/s \cite{Steigman:2012nb} appropriate 
for 33 MeV DM, we find that $\lambda_1 = 3\times 10^{-7}$ or 
$\lambda_2 = 7\times 10^{-7}$ to get the correct relic density.  

In principle, these
couplings are calculable in terms of the U(1)$'$ charge $g'$ and masses of the dark quarks,
and the confinement scale  $\Lambda$ of the dark SU($N$) gauge group.  Such a
calculation is difficult since it requires running down from the fundamental theory 
to scales below $\Lambda$.  It would be interesting to know if such small values of
$\lambda_{1,2}$ are consistent with reasonable choices of the parameters of the 
fundamental theory.  In chiral perturbation theory, these couplings vanish at leading 
order since the pions (according to our assumption of same-charge quarks) are neutral
under the U(1)$'$.  
Naively one expects couplings of the order $\lambda_i\sim
g'^2/(16\pi^2)$ (in analogy with the anomalous vertex of $\pi^0$ to two photons in the
visible sector), implying a rather small value for the coupling strength $\alpha' = 
g'^2/(4\pi) \sim 10^{-5}$. 

\begin{figure}[t]
\centerline{\includegraphics[width=\columnwidth]{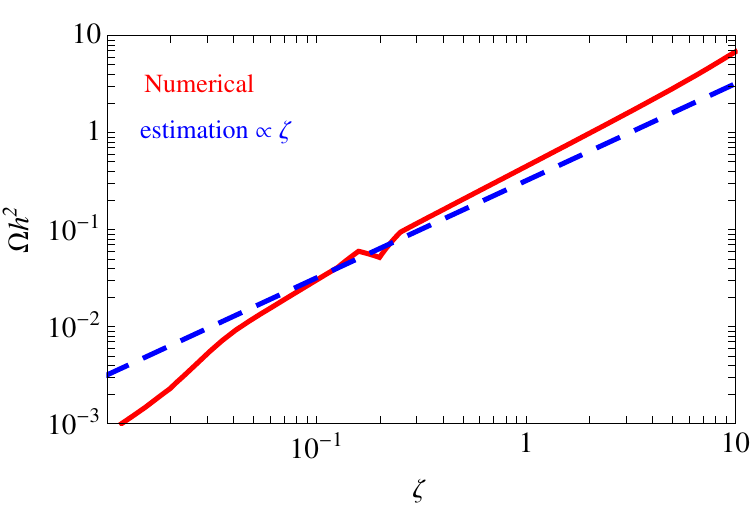}}
\caption{Dependence of relic density of pionic dark matter on
ratio of hidden to visible sector temperatures, $\zeta=T_d/T_\gamma$.
Dashed line shows approximate linear dependence.
}
\label{zeta-omega}
\end{figure}

The preceding calculation implicitly assumes thermal equilibrium between the dark and
visible sectors, but this need not be the case.  In general one expects that the  dark
photons (which we will argue presently should be massless) have a lower temperature
$T_d$ than the visible ones, characterized by a parameter $\zeta = T_d/T_\gamma$.
In that case, $\langle\sigma v\rangle$ must be decreased by a factor $\sim\zeta$ in
order to maintain the correct relic density.  The modification to
the Boltzmann equation in this case has been worked out in ref.\
\cite{Feng:2008mu} (see also 
\cite{Das:2010ts}).  Solving it numerically we find the dependence
of the relic density on $\zeta$ shown in fig.\ \ref{zeta-omega}.
The CMB and BBN give bounds on $\zeta$ in terms of the number of
the number of effective neutrino species, $\Delta N_{\rm eff} \sim
(8/7)\zeta^4$.  Current bounds are roughly consistent with 
$\Delta N_{\rm eff} \lesssim 1$ \cite{Steigman:2012ve,Ade:2013zuv},
hence $\zeta\lesssim 1$.  This bound is quoted in terms of the value 
of $\zeta$ at the time of BBN.  Since photons get heated relative to
dark photons afterwards, the bound on the current value of $\zeta$
is $\zeta_0 < 0.75$ \cite{CyrRacine:2012fz}.

\subsection{CMB and charged relic constraints}

In the preceding computation of the dark pion relic density, 
we also implicitly assumed that the dark photon temperature is not
significantly increased by the dark matter annihilations themselves. 
If the dark sector consists only of the pions and the dark photons,
having no interactions with the standard model, this will not be a
valid assumption and the freezeout calculation must be revisited to
take into account the heating of the dark photons. However
interactions with the standard model will generically be induced via
kinetic mixing $\tilde\epsilon F^{\mu\nu} Z'_{\mu\nu}$ between the
$Z'$ and weak hypercharge.  This would help to maintain thermal
equilibrium of the dark sector particles, but it also introduces a new
problem by  allowing the light $Z'$ to decay into leptons and charged
(visible sector) pions.  This is strongly ruled out by CMB bounds
since the DM is so light \cite{Finkbeiner:2011dx,Cline:2013fm}.  

To avoid this problem, one can take the $Z'$ to be massless.  In this
case, there is no unique way of diagonalizing the gauge boson kinetic
term to remove the mixing.  A convenient choice is that where the
field identified as the photon remains uncoupled to the dark sector,
but the dark particles with U(1)$'$ charge $g'$ acquire an electric
millicharge given by $\epsilon = \tilde\epsilon g'/e$ \cite{Cline:2012is}.  In this
case there is no constraint from injecting electrons into the CMB;
however a fraction $\epsilon^2\alpha/\alpha'$ of annihilations will
produce a visible photon.  The cross section for this process is
constrained as $\langle\sigma v\rangle < 1\times 10^{-27}(m_\pi/{\rm
GeV})$\,cm$^3$/s \cite{Lopez-Honorez:2013cua}, taking into account a
factor of 2 for producing only a single photon.  Comparing to 
$\epsilon^2\alpha/\alpha'$ times the required thermal relic cross
section, we obtain the bound
\be
	\epsilon < 1.7\times10^{-3}\left({\alpha'\over
10^{-5}}\right)^{1/2}\left(m_\pi\over 100{\rm\ MeV}\right)^{1/2}
\label{cmb_bound}
\ee

In addition there are constraints arising from the presence of stable
millicharged relics, the ``baryons'' of the dark sector.   Unless
these have a large relic abundance due to an asymmetry between
particles and antiparticles, their abundance will be highly suppressed
by their strong annihilation cross section.  The abundance  of normal
baryons would be of order $10^{-19}$ in the absence of the baryon
asymmetry \cite{Kolb:1990vq}, and even smaller in the present theory
where the pion mass scale is lower and the nucleon annihilation cross
section is thus larger. 

The cosmological constraints on the kinetic mixing parameter 
of such a small population of millicharged dark baryons are weaker than
the CMB bound (\ref{cmb_bound}).  These are summarized in ref.\ 
\cite{Davidson:2000hf}, and depend upon the mass of the baryon, which
in analogy to QCD we expect to be of order $7 m_\pi\sim 200-700$ MeV.
In this mass region, ref.\ \cite{Davidson:2000hf} shows that the strongest
limit $\epsilon < 0.01$ comes from accelerator experiments.  Stronger
constraints based upon getting too large relic density do not apply to 
this model, since it has a large hadronic annihilation cross section not
assumed in \cite{Davidson:2000hf}.

\section{Dark ``baryons''}
\label{sec:baryon}

We turn to our next example of composite strongly interacting  dark
matter, in which the candidate is analogous to nucleons of the visible
sector: bound states of hidden quarks confined by an unbroken SU(N)
gauge symmetry.  For simplicity we will assume  a common quark mass
$m_q$ and take the number of colors and  light flavors each to be
three as in QCD.  The quark mass  and  confinement scale $\Lambda$ are
considered as the relevant free parameters.   Equivalently, one can
take the pion mass $m_\pi\sim \sqrt{m_q\Lambda}$ and $\Lambda$
as the two free parameters.  Also for simplicity, we will at first
neglect any additional U(1) interactions (dark photons) of the hidden
quarks.

\begin{figure}[t]
\centerline{\includegraphics[width=0.9\columnwidth]{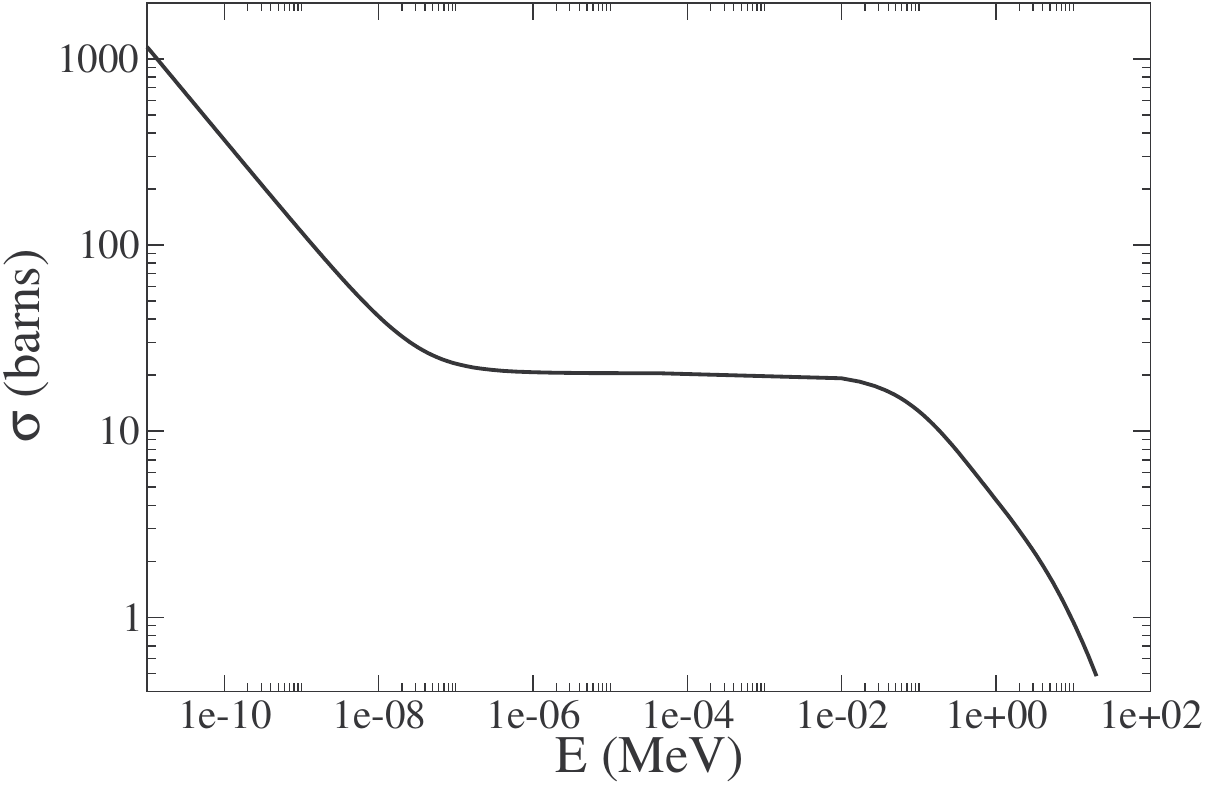}}
\caption{Isospin-averaged elastic cross section for
neutron-proton scattering versus energy, using data from the
ENDF library \cite{endf}.}
\label{np-data}
\end{figure}

\subsection{SIDM constraints}

As a starting point to understand the elastic scattering of dark 
``baryons,''  we consider the  example of real neutron-proton
scattering, whose cross section as a function of center of mass
energy  is shown in fig.\ \ref{np-data}.  To focus on the contribution
from the strong force, we are interested in the flat region starting
at energies above $E_0\sim 0.1$ eV, since the rising cross section
below this value is due to electromagnetic charge-dipole scattering. 
In the plateau, $\sigma \cong 20$ b, so that $\sigma/m \cong 10$
cm$^2$/g, which is 17 times larger than needed for SIDM.  We are
interested to know how the parameters of QCD would need to be rescaled
in a dark analog theory to bring this down to the desired value.

There are two parameters that primarily control the nucleon-nucleon
elastic  cross section.  One is the confinement scale $\Lambda$ of the
strong SU($N$) interactions.  Naively, one would estimate on
dimensional grounds that  the nucleon mass is $N_c \Lambda$ (assuming
current quark masses $m_q \leq \Lambda$), while the cross-section is
$\sigma \sim 4\pi\Lambda^{-2}$.  Therefore
${\sigma/ m} \sim 4\pi/(N_c\Lambda^3)$.  Using this estimate and
the parameters of real-world QCD, $\Lambda\sim 250$ MeV and
$N_c = 3$, we would estimate
$\sigma/m \sim 0.2{\rm\ cm}^2/{\rm g}$.
The naive estimate is too low by a factor of 50.
The origin of the discrepancy is well-known: there is a resonant
enhancement of the cross section due to the weakly bound deuteron.
A better estimate for $\sigma/m$ is given by $2\pi/(N_c\Lambda^2
E_b)$, where $E_b=2.2$ MeV is the binding energy of the deuteron,
which could be considered as the other parameter controlling $\sigma$.

\begin{figure}[t]
\centerline{\includegraphics[width=0.9\columnwidth]{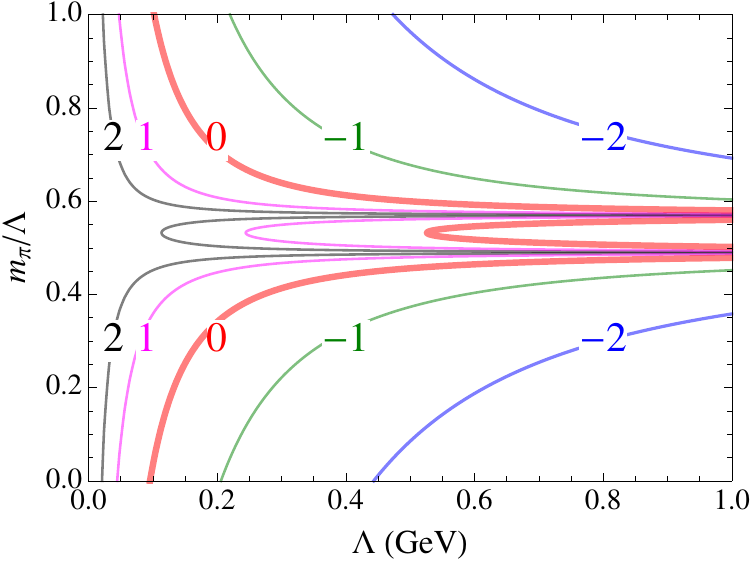}}
\caption{Contours of $\log_{10}([\sigma/m]/[0.6{\rm cm}^2$/g]) in the
plane of confinement scale $\Lambda$ and pion mass $m_\pi/\Lambda$.
Solid line (labeled ``0'') corresponds to desired value.}
\label{baryon}
\end{figure}

Of course $E_b$ is not a fundamental parameter of the theory, but it
gives a clear picture of the physics controlling $\sigma$.  It turns
out that $E_b$ depends sensitively on the mass of the pion (hence
the quark masses).  The effective range parameters for
nucleon-nucleon scattering have been studied as a function of $m_\pi$
in lattice gauge theory \cite{Chen:2010yt}.  They are the scattering
length $a$ and effective range $r_0$ in terms of which the scattering
amplitude is given by
\be
	{\cal A} = {4\pi\over m_N(-ip -a^{-1} + \frac12r_0 p^2 +O(p^4)) }
\ee
where $m_N$ is the nucleon mass and  $p$ is the center-of-mass momentum.

Fitting to the results of fig.\ 1 of \cite{Chen:2010yt}, we can
express the scattering lengths in the singlet and triplet spin
channels as
\be
	a_s = {0.58\, \Lambda^{-1}\over m_\pi/\Lambda - 0.57},\qquad
	a_t = {0.39\, \Lambda^{-1}\over m_\pi/\Lambda - 0.49}
\label{asat}
\ee
where we have taken $\Lambda = 250$ MeV for QCD.  Here $0.49$ and
$0.57$ are the pion-to-$\Lambda$ ratios where the deuteron and
the dineutron become bound; they are unbound for lighter pions and
bound for heavier pions.  In the analysis of
\cite{Chen:2010yt}, only $m_\pi$ was varied while $\Lambda$ was held
fixed, but on dimensional grounds, eq.\ (\ref{asat}) should encode the
right dependence if $\Lambda$ were to be varied.  We can therefore
predict the average scattering cross section for low-velocity 
nucleons  in a dark
sector similar to QCD, but with  different confinement scale and 
light quark masses,
\be
	\sigma = \pi(a_s^2 + 3 a_t^2)
\ee
and find values of $\Lambda$ and $m_\pi$ in a dark analog of QCD
that would give the desired value
of $\sigma/m_N$, using $m_N = 3.8\,\Lambda$ to agree with the visible
nucleon mass.

The results are plotted in fig.\ \ref{baryon},
which shows contours of $\log_{10}([\sigma/m]/[0.6\,{\rm cm}^2$/g]) 
as a function of $\Lambda$ and $m_\pi/\Lambda$.  The values
$m_\pi/\Lambda = 0.49$ and $0.57$ are where the triplet and singlet
scattering lengths diverge, respectively.  For  
$\Lambda \gtrsim 1$ GeV, $m_\pi/\Lambda$ needs to be close to these special values
to have a large enough cross section, but for 
$\Lambda \lesssim 1$ GeV, this tuning is not necessary.  For $\Lambda <
100$ MeV, large values of the pion mass $m_\pi>\Lambda$ would be
required to keep the cross section sufficiently low.

We must still verify that the energies of interest for dark matter
scattering coincide with the flat region of the cross section.
Since we do not require dark photons in this scenario, the rising
part below 0.1 eV in fig.\ \ref{np-data} is not present in the dark
analog.  The fall-off after the plateau occurs when the inverse
momentum of the nucleons starts to exceed the length scale
$\sqrt\sigma \sim 3.5\,a$, where 
in the plateau region, $a=\sqrt{\sigma/4\pi}=13$ fm. This corresponds
to a center-of-mass energy $p^2/m_N = (4\pi a^2 m_N)^{-1} = 0.02$ MeV,
which agrees with fig.\ \ref{np-data}.  For the dark baryons
to be SIDM, we thus require that $(\sigma m_N)^{-1} > m_N v^2$ up to
velocities $v\sim 100$ km/s.  This can be written as the constraint
$m_N < (v^2\sigma/m_N)^{-1/3} = 15$ GeV, hence $\Lambda < 4$ GeV
which is consistent with the parameter space plotted in fig.\
\ref{baryon}.

It has been pointed out that, if asymmetric fermionic dark matter has
strong attractive interactions, their accumulation in neutron stars
leads to a compact bound state that can cause gravitational collapse
of the star \cite{Kouvaris:2011gb,Bramante:2013nma}, yielding tighter constraints than
from halo ellipticity or the Bullet Cluster. These considerations
however do not apply to composite models such as the present one. 
Like the neutrons and protons making up neutron stars, dark nucleons
are expected to exhibit short-range hard-core repulsion due to the
degeneracy pressure of the underlying dark quarks, leading to an
equation of state qualitatively similar to that for neutron star
matter.  So while attractive interactions could form a dark nucleus,
the mass of dark matter required to achieve gravitational collapse
should be comparable to the mass of the neutron star itself, safely
more than the amount that is expected to accrete in a neutron star.

\subsection{Dark baryon relic density}

Like their visible counterparts, the dark nucleons have a conserved
number, and so must be asymmetric dark matter.  We do not attempt to
explain the origin of the asymmetry here (indeed that of the visible
baryons is still unknown); it presumably arises from physics at much
higher scales than that of the dark matter, which we have determined
to be of order $(0.1-1$ GeV).  However one may wonder why in this case
the dark pions that are necessarily present have a small enough relic
density.  A simple possibility is that the quarks are massless so that
the pions are true Goldstone bosons, and contribute only to the dark
radiation density of the universe.  Fig.\ \ref{baryon} shows that 
$m_\pi=0$ is compatible with $\Lambda\cong 100$ MeV.

If the pions are massive, the existence of massless dark photons
coupling to the hidden quarks would allow for them to efficiently
annihilate, but this also provides a new long-range interaction
between the dark nucleons, which could complicate its viability as a
dark matter candidate, and require additional species to maintain the
U(1)$'$ charge neutrality of the universe.  A less problematic
alternative would be to give the dark photons ($Z'$) masses greater than
$2 m_e$, such that $\pi\pi\to Z'Z'$ annihilation would still be
efficient, while $Z'\to e^+e^-$ through kinetic mixing of 
$Z'$ to the photon would allow the $Z'$s to decay.  We consider
the possibility of unstable dark pions below.

\subsection{Interactions with the standard model}

\begin{figure}[t]
\centerline{\includegraphics[width=\columnwidth]{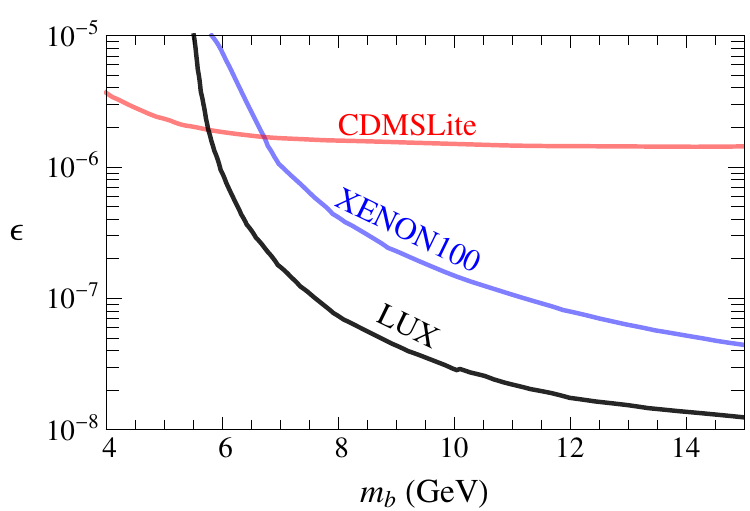}}
\caption{Constraints on kinetic mixing of massive $Z'$ and photon
from scattering of dark baryons on protons, as a function of 
dark baryon mass, 
from the LUX \cite{Akerib:2013tjd}, XENON100 
\cite{Aprile:2012nq} and
CDMSlite \cite{Agnese:2013jaa} experiments, assuming $g'=1$ and $m_{Z'}=1$ GeV.  The bound
on $\epsilon$ scales as $m_{Z'}^2/g'$ for other values of $g'$ 
and $m_{Z'}$.}
\label{baryon-dd}
\end{figure}

An interesting consequence of coupling the dark quarks to a massive
$Z'$ is that kinetic mixing of the $Z'$ and the photon would allow
for scattering of dark baryons on protons, hence a channel for
direct detection.  After diagonalizing the gauge boson kinetic matrix,
the proton acquires a dark millicharge $\epsilon e$; see for example
\cite{Chen:2009ab}.  Assuming the dark baryon has U(1)$'$ charge
$3g'$ (taking the charge to be $g'$ for each of three hidden sector
quarks),  the cross
section for proton-baryon scattering is
\be
	\sigma_{pb} = 144 \pi\, \alpha\, \alpha' \epsilon^2{\mu^2\over
m_{Z'}^4}
\ee
where $\mu = m_b m_p/(m_b+m_p)$ is the dark baryon-proton reduced
mass.
The resulting constraints on the kinetic mixing parameter
$\epsilon$ from the LUX \cite{Akerib:2013tjd}, XENON100 \cite{Aprile:2012nq} and CDMS\-lite 
\cite{Agnese:2013jaa} experiments are shown
in fig.\ \ref{baryon-dd}.

Another possibility is to imagine heavy mediators producing
isospin-violating 
dimension-6 couplings between the dark quarks ${\bf q}_i$
(where $i$ is the flavor index) and light
standard model particles such as the electron:
\be
	c_{ij}\Lambda_h^{-2}(\bar{\bf q}_i\gamma_5\gamma_\mu{\bf q}_j)\, 
	(\bar e\gamma_5\gamma^\mu e)
\label{dim6}
\ee
with $c_{ij}$ being coefficients of order 1.
We choose $\gamma_5$ couplings to match the parity of the
pion; this operator allows for the decays $\pi\to e^+e^-$.
By estimating the decay rate of the pion as $\Gamma_\pi = n\sigma v$
where $\sigma$ is the ${\bf q}_i{\bf\bar q}_j\to e^+e^-$ cross section from
(\ref{dim6}), $n\sim \Lambda^3$ is the density of quarks in the
pion, and $v\sim 1$, we obtain
\be
	\Gamma_\pi \sim {m_\pi^2\Lambda^3\over 12\pi\Lambda_h^4}
\label{gammapi}
\ee
To avoid overclosing the universe around the time of big bang
nucleosynthesis (BBN), $\Gamma_\pi$ should be greater than 
$\sim 1$ s$^{-1}$.  This gives an upper bound on the heavy physics
scale
\be
	\Lambda_h < 8\,{\rm TeV}\left(m_\pi\over 10{\rm\
MeV}\right)^{1/2}\left(\Lambda\over 100{\rm\ MeV}\right)^{3/4}
\label{Lhbound}
\ee  

Interestingly, the same operator allows for scattering of the 
dark baryons from electrons.  
Recently the first experimental constraints on dark matter
scattering from electrons were published \cite{Essig:2012yx},
giving limits from $3\times 10^{-38}$ cm$^2$ to  
$2\times 10^{-37}$ cm$^2$ for dark matter of mass 0.1 to 1 GeV.
The cross section for electron scattering with baryons containing
$N$ quarks is $\sigma_{eb} = (3N/\pi)\mu^2/\Lambda_h^4$, where 
$\mu\cong m_e$ is the electron-baryon reduced mass.  With $N=3$
we find the 
limit $\Lambda_h > 10$ TeV.  This bound starts to conflict with the 
need for pions to decay before big bang nucleosynthesis if $m_\pi$ and
$\Lambda$ are near the lowest values indicated on fig.\ \ref{baryon},
as the example of $m_\pi = 10$ MeV, $\Lambda = 100$ MeV
shows in eq.\ (\ref{Lhbound}).

\section{Dark glueballs}
\label{gball}

If the quarks of the hidden SU(N) are sufficiently heavy, 
then the lightest stable particle is the glueball
$\phi$, whose self-interaction cross section and mass can be
estimated as $\sigma \sim 4\pi/\Lambda^2$, $m_\phi\sim 5.5\,\Lambda$.
(We use the example of QCD where a likely glueball candidate
has mass 1370 MeV \cite{Ochs:2013gi} to get the factor of 5.5.)  For 
SIDM,
this leads to the requirement $\Lambda\cong 90$ MeV, 
$m_\phi\cong 500$ MeV, hence the dark quark mass should obey
$m_q \gtrsim 250$ MeV in this scenario. Like for
baryons, we expect the cross section to be velocity independent for
c.m.\  energies $E < (\sigma m_\phi)^{-1}$.  This 
requires $m_\phi < 15$ GeV, which thus imposes no additional
constraint.

It is challenging (perhaps impossible) to design a mediator that allows for thermal
freezeout of dark glueballs by annihilation into lighter particles.
Unlike pions or nucleons, whose stability could be ensured by unbroken
isospin or baryon number, nothing forbids glueballs from decaying into
the lighter particles once any mediator is introduced.  We do not try
to explain the relic density of glueballs here.   It could arise from
initial conditions---the relative efficiency of inflationary reheating
of the visible and hidden sectors---as long as the reheating
temperature was too low to bring the two sectors into equilibrium
at early times.

\subsection{CMB constraint versus direct detection}

Nevertheless there are some generic statements that can be made
about the nature of mediators between dark glueballs and the standard
model.  Suppose that new particles at the high scale $\Lambda_h$ 
induce an
effective interaction between the dark gluon (with field strength
$G_{\mu\nu}$) and standard model gauge singlet operators ${\cal O}_{\rm
sm}$ of dimension $n$:
\be
	{1\over\Lambda_h^n}\,G_{\mu\nu}G^{\mu\nu}\, {\cal O}_{\rm sm}
\label{effop}
\ee
We assume that the gluon operator $G^2$ interpolates between $\phi$ and the vacuum
as $\langle 0|G^2|\phi\rangle\sim (m_\phi\Lambda)^{3/2}$; this
parametrization agrees with the $\Gamma\sim n\sigma v$ estimate used 
previously for pion decays, with $n\sim\Lambda^3$.
Thus (\ref{effop}) leads to 
decays of $\phi$ if ${\cal O}_{\rm sm}$ consists of
states that are lighter than $m_\phi$, such as $\gamma$, $e$, $\mu$.
Such decays are subject to stringent CMB constraints; for example
the lifetime for  $\phi\to e^+e^-$ 
for 500 MeV dark matter must satisfy
$\tau > 4\times 10^{24}$ s $\cong 10^{49}$ GeV$^{-1}$ 
\cite{Diamanti:2013bia}. Assuming that 
$\langle e\bar e|{\cal O}_{\rm sm}|0\rangle\sim m_e m_\phi^{n-3}$
(since the operator is chirality-suppressed and the electrons have energy of order $m_\phi$), we find the decay rate $\Gamma_\phi\sim
(m_\phi/16\pi)(\Lambda/m_\phi)^3(m_e/m_\phi)^2(m_\phi/\Lambda_h)^{2n}$, hence the constraint
\be
	\left({\Lambda_h\over m_\phi}\right)^{n} \gtrsim 10^{19}
\label{lhbound}
\ee
taking $m_\phi = 5.5\Lambda = 0.5$ GeV.
For a 4-body decay such as $\phi\to e^+e^-e^+e^-$ this would be 
increased by $(m_\phi/m_e)/(4\pi^2)\sim 10^{3/2}$ since it need no longer be
chirality suppressed, but does suffer from 
a phase space reduction $\sim (4\pi^2)$.\footnote{
The ratio of phase spaces for the 4- and 2-body rates is of order
$(m_\phi^2/4\pi^2)^2$.  We have extracted the factors of $m_\phi$ to make the
dimensionless ratio in (\ref{lhbound}) and taken the square root since
(\ref{lhbound}) is a bound on the amplitude.}

The same interactions that cause 2-body glueball decays give rise to
elastic scattering with visible matter, because
the $GG$ operator also interpolates between the vacuum
and the two-glueball state, with $\langle 0|GG|\phi\phi\rangle \sim
m_\phi^2$. (For this part of the argument, the distinction between
$m_\phi$ and $\Lambda$ is not important.)
 The cross section for $\phi e\to\phi e$ from
(\ref{effop}) is therefore of order
\be
	\sigma_{\phi e} \sim m_\phi^{-2}
	\left({m_\phi\over \Lambda_h}\right)^{2n}\lesssim
	10^{-66}{\rm\ cm}^2
\label{sbound}
\ee
using (\ref{lhbound}).  This is many orders of magnitude below the current
limit discussed below eq.\ (\ref{Lhbound}).
 We see that the CMB bound on $\phi$ decays
is so strong that crossing symmetry implies that its scattering interactions are 
necessarily negligible.

In passing we observe that a similar argument shows that one cannot
get a large enough annihilation cross section of glueballs for them to
have the right thermal relic density if their lifetime is greater than
that of the universe.   Consider for example the coupling of glueballs
to light $Z'$ gauge bosons through the operator
$\Lambda_h^{-4}\tr(G_{\mu\nu}G^{\mu\nu})
\,Z_{\alpha\beta}'{Z'}^{\alpha\beta}$, which gives rise to both 
decays $\phi\to Z'Z'$ and annihilations $\phi\phi\to Z'Z'$. Demanding
that the lifetime exceed $10^{18}$ s gives a cross section less than
$10^{-66}$ cm$^2$.

\begin{figure}[t]
\centerline{\includegraphics[width=\columnwidth]{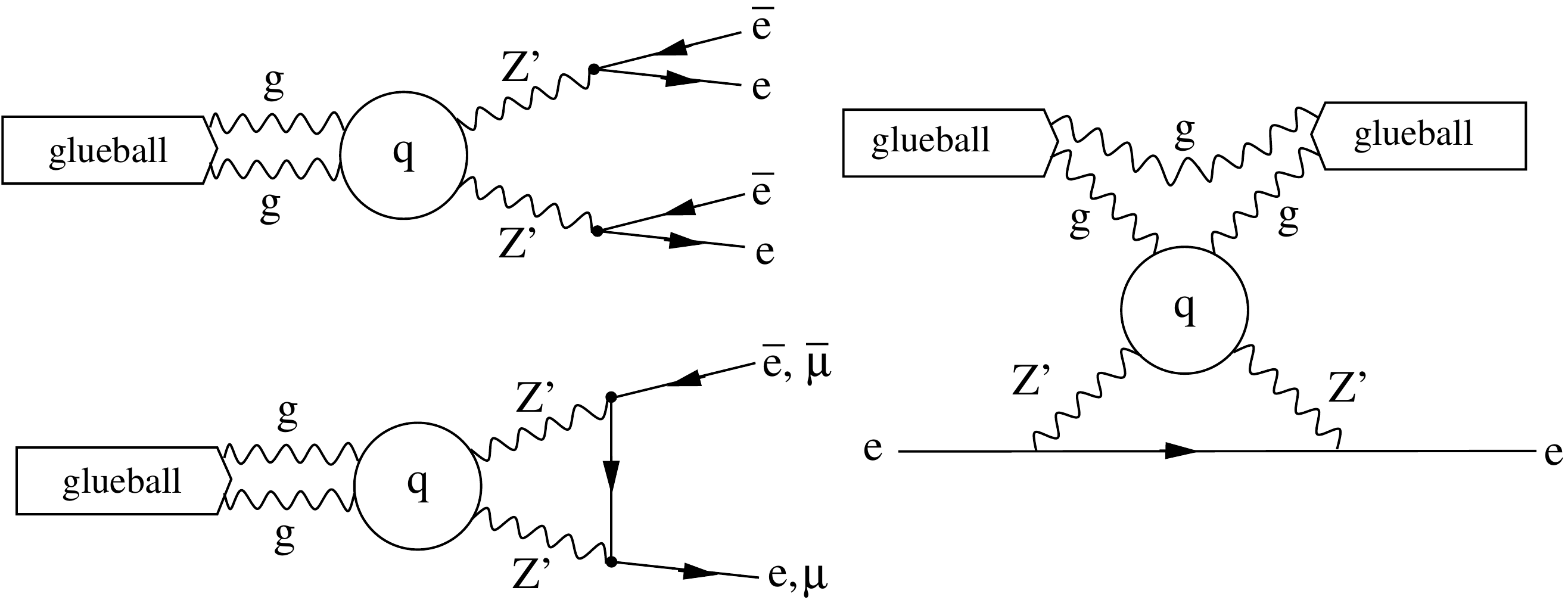}}
\caption{Upper left: decay of dark glueball into $e^+e^-e^+e^-$
by virtual $Z'$ emission through dark quark loop.
Lower left: related decays into $e^+e^-$, or 
$\mu^+\mu^-$, which is subdominant
if $m_q<8$ GeV.
Right: leading contribution to dark glueball-electron scattering,
which is shown to be negligible.}
\label{gbdecay}
\end{figure}

\subsection{Heavy $Z'$ mediator between glueball and SM}

As a concrete example, we consider as mediator a heavy $Z'$ gauge
boson that couples to the hidden quarks and to leptons
with strength $\alpha'$.  Integrating out the hidden quarks and
$Z'$s leads to several possible operators giving glueball decay,
including ${\cal
O}_1 = (\bar e\gamma^\mu e)^2$ and ${\cal O}_2 = \bar \mu  \mu$. 
(We will presently see that the operator $\bar e e$ gives rise to a
much smaller contribution to the decay width, hence we focus on
muons for ${\cal O}_2$.)  Fig.\
\ref{gbdecay} shows the corresponding diagrams for  decays into
$e^+e^-e^+e^-$ and $\mu^+\mu^-$.  The first one has heavy scale given by
$\Lambda_1^{-6}\sim 64\pi\alpha_N\alpha'^2 m_\phi^2/(360\, m_q^4\,
m_{Z'}^4)$  where $\alpha_N=g_N^2/4\pi$ is the hidden SU(N)  gauge
coupling.\footnote{
The quark loop in the 4-body decay diagram generates an
Euler-Heisenberg-like effective interaction between the gluons and the
$Z'$ field strength, $\alpha_N\alpha'/(360\, m_q^4)[\tr(G^2)Z'^2 +
7\,\tr(G\tilde G)Z\tilde Z]$.  The factor of $64\pi$ comes from $g'^2=4\pi\alpha'$,
the coefficients $1+7=8$, 
each $Z'$ field strength containing two vector fields, and the 
$Z'$ momenta going as $p_1\cdot p_2\sim m_\phi^2/2$.  To estimate the 
second decay diagram, it is easiest to first do the loop containing the 
$Z'$s, which is dominated by momenta of order $m_{Z'}$ and
requires mass insertions of the SM fermion and the dark quark.  This loop is of order
$g'^4 m_\mu m_q/(16\pi^2\, m_{Z'}^4)$.  The quark loop, dominated by momenta
of order $m_q$, now has only three
propagators, requiring one more mass
insertion, and so contributes $\sim g_N^2 m_q/(4\cdot16\pi^2\, m_q^2)$ to the $G_{\mu\nu}
G^{\mu\nu}$ effective operator.
}
The second operator is chirality-  and loop-suppressed and has
$\Lambda_2^{-3} \sim m_\mu \alpha_N \alpha'^2/(16\pi m_{Z'}^4)$.
Taking into account the 
phase space ratio $\sim (m_\phi^2/4\pi^2)^2$, the 2-body and 4-body
decay rates become comparable for  $m_q\cong 0.7$ GeV. 
This is somewhat larger than the minimum dark quark mass of 
$m_\phi/2\sim 0.25$ GeV needed to ensure that the glueball is lighter
than the meson. The CMB constraint (\ref{lhbound})  then leads to the bound
\be
	m_{Z'}\gtrsim 2.3\,{\rm TeV} \left(\alpha_N\alpha'^2\over
10^{-5}\right)^{1/4}\left\{\begin{array}{ll}
	x^{-1}, & x<1\\
	1,&x>1\end{array}\right.
\label{Zpconst}
\ee
where $x=m_q/(0.7\,{\rm GeV})$. The strong coupling $\alpha_N$
should be evaluated at the scale $m_q$, hence $\alpha_N\sim 1$
if $m_q$ is near the confinement scale, but smaller otherwise.

This shows that the new physics can be at a relatively low
scale accessible at LHC, despite the large ratio in (\ref{lhbound}).
For a $Z'$ with couplings  $\alpha'\sim 0.003$  to ordinary quarks 
(the ``sequential standard
model'', SSM), ATLAS obtains the limit 
$m_{Z'} \gtrsim 3$ TeV \cite{ATLAS:2013jma} from 
resonant dilepton searches.  Thus it is possible that a mediator between
the standard model and metastable glueball dark matter, 
consistent with CMB constraints on the glueball decays,
could be discovered at the LHC. 

\begin{figure}[t]
\centerline{\includegraphics[width=\columnwidth]{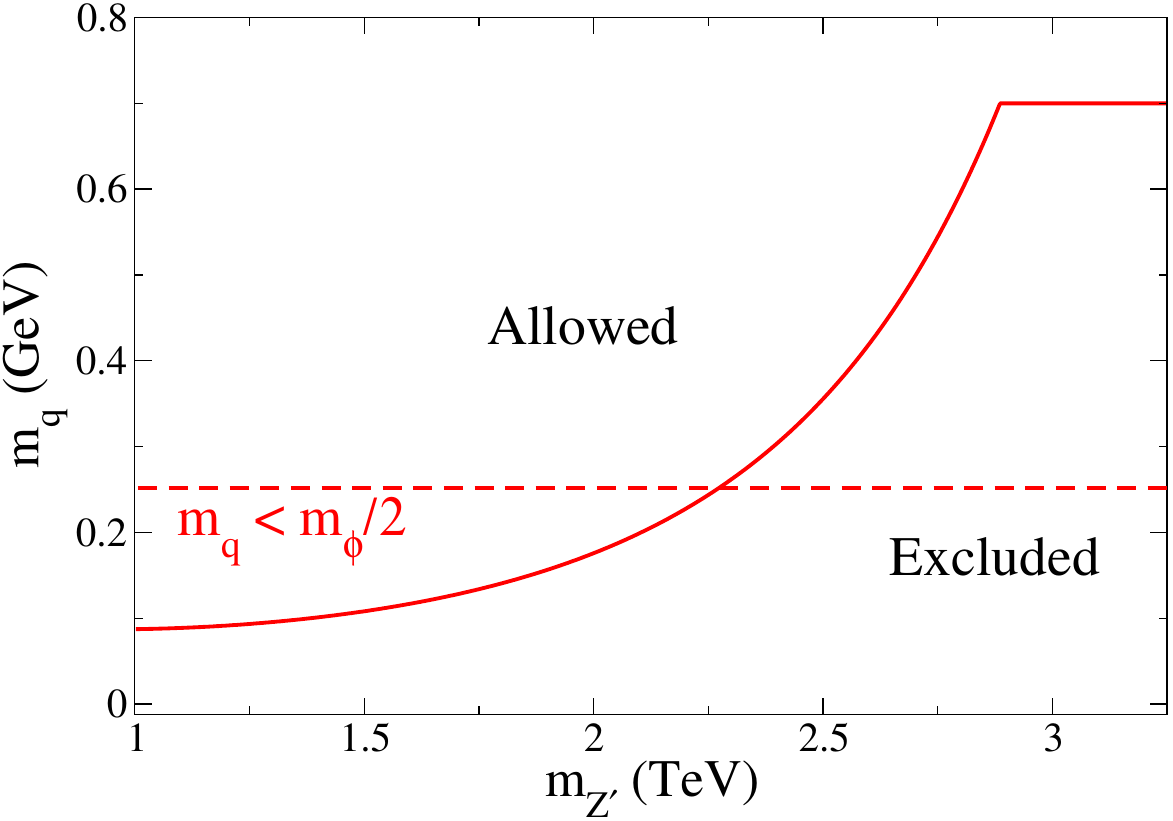}}
\caption{Lower bound from CMB on dark quark mass $m_q$ as a function of
$m_{Z'}$, for couplings $\alpha'$ that saturate the ATLAS constraint
from dileptonic decays of $Z'$ \cite{ATLAS:2013jma}.  Below the dashed line
is theoretically disfavored since glueballs would be heavier than mesons
in that region.}
\label{dilepton}
\end{figure}

To extract limits from the ATLAS results, 
we computed the expected number of dilepton events in the model
with coupling 
$\sqrt{4 \pi \alpha' } Z'_\mu \bar f \gamma^\mu f$ to all SM fermions
and dark quarks, but ignoring the contribution of $Z'$ decays into
dark quarks to compute 
branching ratio $B$ to leptons. (This is justified if the 
number of dark quarks is small
compared to the number of SM fermions).
Comparing to the limit on $\sigma B$ of ref.\ \cite{ATLAS:2013jma},
we obtain the constraint as a function of $Z'$ mass 
\be
	\log_{10}\alpha' < -5.71 + 0.410\, y + 0.267\, y^2
\label{alphaconst}
\ee
where $y=m_{Z'}$ in TeV.  To compare this with the CMB bound
we treat (\ref{alphaconst}) as an equality
 to eliminate $\alpha'$ in (\ref{Zpconst}), and assume $\alpha_N\sim 1$, resulting in 
a lower bound on $m_q$ as a function of $m_{Z'}$ shown in 
fig.\ \ref{dilepton}.  This is the CMB bound on models that are 
discoverable in the next run of the LHC.  The range of allowed
quark masses is mostly consistent with our requirement that glueballs be
lighter than mesons in the dark sector for this scenario (indicated by the
dashed line), while still being
relatively small.  
The implication is that models on the verge of discovery through $Z'$
production at the LHC could also be close to having an impact on the CMB
for reasonable choices of parameters.  Such a $Z'$ might also reveal the
number of dark quark species through the measurement of its invisible width.

\subsection{Higgs portal mediation}

As a second example, we imagine that the heavy particles are scalars
$S$ in the fundamental representation of the hidden SU(N), that 
communicate with the standard model through the Higgs portal
interaction $\lambda|S|^2|H|^2$.  Integrating out the scalar gives
the SM operator ${\cal O} = |H|^2$ with $\Lambda_h^{-2}
= \lambda \alpha_N / m_S^2$.  The Higgs boson mediates the decay
$\phi\to\bar\mu\mu$, with rate 
\be
	\Gamma_\phi \sim {m_\mu^2(m_\phi\Lambda)^3 m_\phi\over
	16\pi \Lambda_h^4\, m_h^4}
\ee
Demanding that $\Gamma < 10^{-49}$ GeV as before, we obtain the bound
\be
	m_S > 10^7{\rm\ 
   GeV}\left(\lambda\alpha_N\over0.01\right)^{1/2}
\ee
which is inaccessible to the LHC.  The bound is much stronger in this
case than for the $Z'$ mediator because the matrix element is
suppressed by only $1/m_S^2$; compare to $1/m_{Z'}^4$ in the previous
model.

\subsection{Neutrino portal mediation}
Since the CMB constraints are so severe, one might ask whether dark
glueballs could have larger interactions with the SM if they decayed
only into neutrinos rather than charged leptons.  However the
constraints from Super-Kamiokande on DM decay into neutrinos are 
still quite strong \cite{Bell:2010fk}: for decay of a 500 MeV
glueball, the lifetime must exceed $2\times 10^{22}$s, which is only
100 times weaker than the CMB bound on leptonic decays.  
To illustrate, we consider an example in which the
SM operator coupling to $G_{\mu\nu}G^{\mu\nu}$ is the neutrino portal
\cite{Falkowski:2009yz}
$(LH)^2$ where $L$ is a charged lepton doublet.  Then the effective
operator is
\be
	{\cal O} = \Lambda_h^{-5}(LH)^2\, G^2 = {\Lambda'_h}^{-3}
	\bar\nu\nu\, GG
\ee
In the second form, we absorb the Higgs VEVs $v^2$ into
$\Lambda_h^{-5}$ to display the relevant form of the operator at 
energies below the weak scale.  The decay rate is then
$\Gamma \sim (\Lambda m_\phi)^3 m_\phi
/16\pi{\Lambda'}_h^{6}$ and we obtain the bound 
$\Lambda'_h > 5600$ TeV.

\begin{figure}[t]
\centerline{\includegraphics[width=0.7\columnwidth]{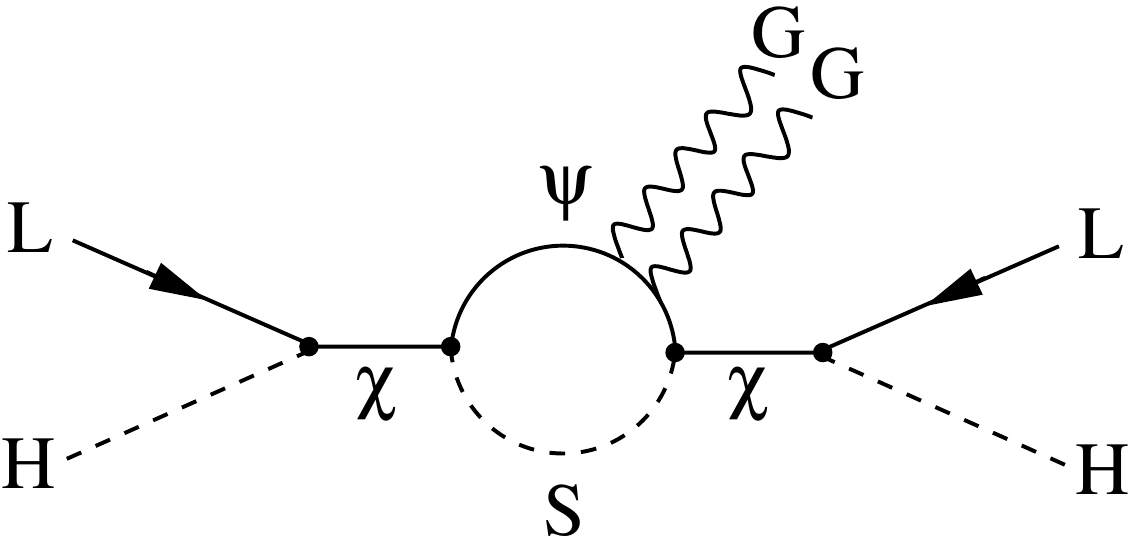}}
\caption{Diagram to generate neutrino portal interaction $(LH)^2$
coupling to the dark glueball operator $G_{\mu\nu}G^{\mu\nu}$.}
\label{n-portal}
\end{figure}

As a concrete example, a model that can generate the desired
interaction was presented in ref.\
\cite{Cline:2012bz}, where the dark sector contains a
scalar $S$ and fermion $\psi$ in the fundamental representation
(here however we take their electric charges to vanish) and a singlet
fermion $\chi$, with Yukawa interactions $y_\chi\bar\chi S^*\psi$ and
$y_\nu\bar\chi H L$.
The $(LH)^2\,G_{\mu\nu}G^{\mu\nu}$ effective operator is generated by
the diagram shown in fig.\ \ref{n-portal}, with coefficient
$\Lambda_h^{-5} = (y_\chi^2 y_\nu^2/16\pi^2)\, m_\chi^{-2}\,  m_{S/\psi}^{-3}$
where we take $m_S\sim m_\psi = m_{S/\psi}$.   There is also a 
seesaw contribution to the neutrino masses of order $(y_\nu
v)^2/m_\chi$, which must be $\lesssim 0.2$ eV.  Regardless of that or
other details of the theory however, the main point is that scattering
of the glueballs with visible matter, mediated by the same diagram 
as in fig.\ \ref{n-portal} (but with gluon lines associated to different
glueballs as in fig.\ \ref{gbdecay}, right), is suppressed by the same
large factor of ${\Lambda'}_h^{6}$ as in the decay rate.
The neutrino portal thus offers no substantial relaxation of the 
scale by which interactions of the glueballs with the visible sector
must be suppressed.

\section{Summary and conclusions}
\label{conclusions}

The possibility that dark matter self-scatters elastically with  a
velocity-independent cross section $\sigma \sim 1{\rm b}\times
(m$/GeV) is motivated by the cusp/core and too-big-to-fail problems of
structure formation with  cold dark matter.  These problems may find
alternative solutions, as we mentioned in section \ref{intro}; in that
case the quoted cross section is at least still allowed by current
constraints.  In previous literature, the possibility to get such a
large cross section from light mediators was explored.  Here we have
investigated for this purpose various forms of composite dark matter,
that can naturally have strong self-interactions.  We examined the cases
where dark matter is analogous to atoms, molecules, mesons,
baryons, or glueballs.

Atomic dark matter bound by a new U(1)$'$ interaction was found to be viable as
a SIDM candidate for a large range of values of the ratio $R=m_\pd/m_\de$ (the
dark proton to electron masses), with U(1)$'$ coupling $\alpha'\gtrsim 0.03$
and mass $m_\dH \sim 0.3 (R/\alpha')^{2/3}$ GeV (this follows from the
rough estimate $\sigma \sim 100 a_0^2$ of the cross section).  The same
estimate is also valid if the atoms are primarily bound into  $\dH_2$
molecules.  In both cases, the dark coupling should satisfy $\alpha' \gtrsim
0.03$ to avoid a significant fraction of  ionized constituents.  The question
of whether dark atoms will exist mostly within molecules is interesting but
beyond the scope of the present paper.  However the absence of dark stars
(hence ionizing dark radiation) suggests that molecules will be prevalent. 
See ref.\ \cite{Cline:2013pca} for a more detailed discussion.
Another interesting feature of dark atoms is that the cross-section typically
has nontrivial velocity dependence at the low velocities relevant to cosmology,
generally being larger at lower velocities relevant in dwarf galaxies than
at higher velocities relevant in clusters.

In the case of dark mesons, bound by an SU(N) interaction with 
confinement scale $\Lambda$,  a low mass $m_\pi\sim 30-100$ MeV is
required for them to be SIDM, depending upon the number of hidden
quark flavors and the ratio $m_q/\Lambda\sim m_\pi^2/\Lambda^2$.  We found that dark pions
in this mass range can have a thermal origin if there is a very weakly
coupled ($\alpha'\sim 10^{-5}$) massless U(1)$'$ in the hidden sector
that kinetically mixes with the photon. In this case the hidden quarks
acquire electric millicharges $\epsilon e$.  As long as the dark
baryons (which are also millicharged) have no asymmetry, their relic
abundance is very small and the constraint on $\epsilon$ from charged
relics is weak, $\epsilon \lesssim 2\times 10^{-3}$. A comparable
and more secure bound $\epsilon \lesssim 2\times 10^{-3}$ arises
from  CMB constraints on annihilations into visible plus dark photons.

If the dark matter is in the form of nucleon-like bound states,
it can be SIDM with 
masses typically in the range $m_B \sim 0.1-1$ GeV.  Larger values 
are possible if the dark pion mass happens to be close to where one of
the scattering lengths diverges, $m_\pi \cong 0.49\,\Lambda$ or
$0.57\,\Lambda$, suggested by lattice studies.  We determined  the
relations between $\Lambda$	 and $m_\pi$ that give the desired
SIDM cross section in QCD-like dark sectors. To avoid overclosure of
the universe by the dark pions, they should either be massless  or
else unstable.  The latter case implies interactions with the 
standard model that can lead to direct detection.

Dark glueballs, whose mass should be $\sim 500$ MeV in order to be
SIDM, are generally more problematic than the other candidates in
terms of having the right relic density and additional direct or
indirect detection signatures.  Their couplings to the standard model
are tightly constrained by their effect on the CMB through decays.  
These interactions are thus shown to be too weak to mediate
direct detection.
However, they can be consistent with mediators at the TeV scale that
could be discovered at the LHC, as we showed in an explicit example
with a $Z'$ mediator.
A low reheat temperature after inflation would be needed to prevent
a thermal population of glueballs, which would overclose the 
universe.

Apart from glueballs, all of the candidates we have considered are
motivated (through relic density considerations) to have significant
interactions with the standard model, including  couplings to a dark
photon, either massless or massive, that can mix with the visible
photon. This gives the possibility for direct or indirect signals that
could provide additional observational probes of the  models.   In the
case of dark atoms, the kinetic mixing parameter $\epsilon$ is already
strongly constrained by the XENON100 bound, as shown in fig.\
\ref{eps-lim}.  A prediction of our dark meson model is the existence
of a very small population of millicharged dark baryons that might be
discovered in more sensitive searches for anomalous isotopes. 
Conversely if dark baryons dominate, then the same interactions that
would deplete dark mesons through decays can generate scattering of
the baryons on electrons at a level that is already constrained using
XENON10 data. 

We expect future simulations of structure formation and 
astrophysical studies \cite{Saxton:2012ja}-\cite{Baushev:2013ida} to improve our
understanding of whether strong self-interactions of dark matter are
really needed, or to what extent they are allowed.   If such
interactions arise from the composite nature of dark matter,
our study shows that its mass should be $\sim 10-100$ GeV if it is
in atomic/molecular form, or $\sim 0.1-1$ GeV if it is 
mesonic/baryonic.  In either case, there are several independent
observables that could provide complementary tests.

\smallskip

{\bf Acknowledgments.}  We thank Carlos Frenk, Annika Peter, 
Francis-Yan Cyr-Racine,
Kris-Sigurdson, Sean Tulin, Alfredo Urbano and Jure
Zupan for helpful discussions.  JC thanks the Aspen Center for Physics
for providing a stimulating working environment while this research
was in progress.
\bigskip

\appendix

\section{BBN constraint on dark photons}
\label{bbnapp}
\begin{figure}[t]
\centerline{\includegraphics[width=\columnwidth]{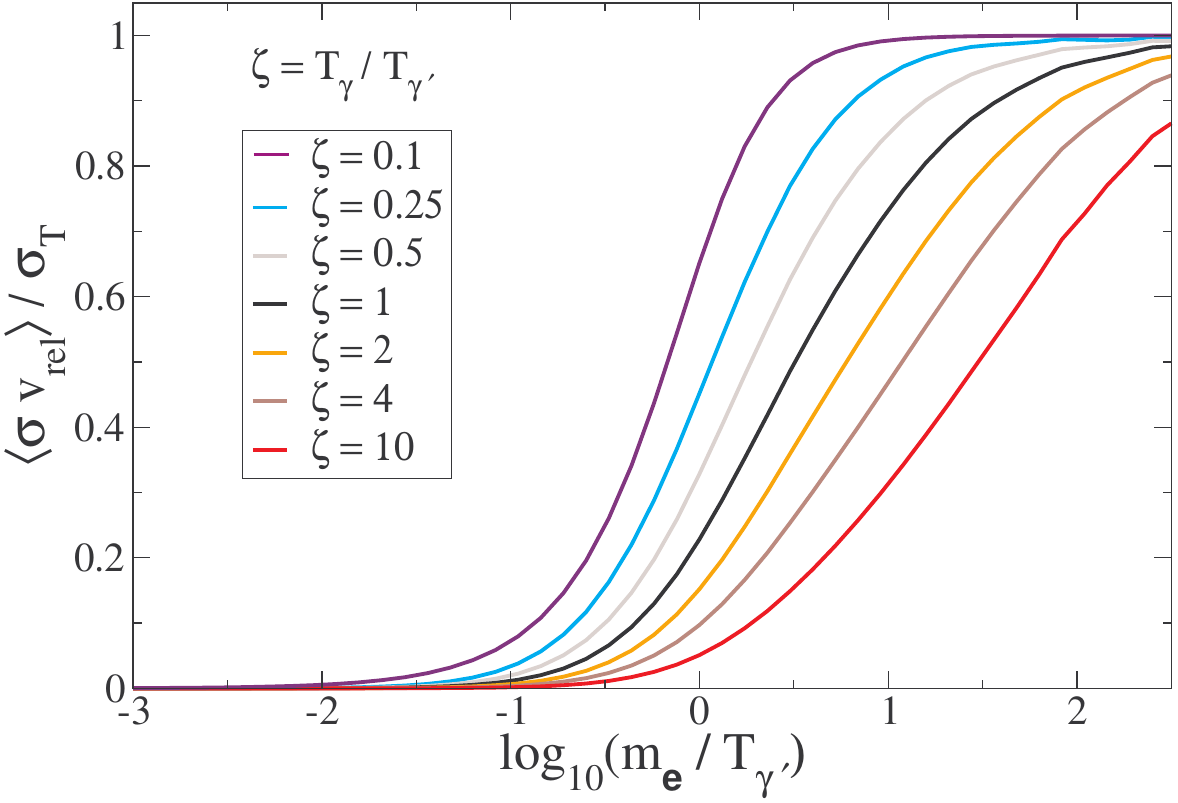}}
\caption{Thermally averaged cross section for 
$\gamma\de\to\gamma'\de$ at several values of $\zeta =
T_\gamma/T_{\gamma'} \equiv T/T_d$, in order of increasing 
$\zeta$ from top to bottom.  The dependence upon $\zeta$
only applies for $\gamma\de\to\gamma'\de$, not the reverse
process, which corresponds to $\zeta=1$.
}
\label{fig:therm}
\end{figure}

In this appendix we provide details of our implementation of the
BBN bound on the dark photon temperature.  In both the approximate
and more exact methods, we make use of the thermally averaged cross
section for mixed Compton scattering on dark electrons, $\gamma'\de\leftrightarrow
\gamma\de$.  In the rest frame of the initial $\de$, the cross section
is given by $\sigma(w) = \sigma_T f(w)$ where $w$
depends upon the initial photon energy $E_i$ as $w\equiv
E_i/m_\de$ and \cite{Itzykson:1980rh} (see eq.\ (5-116))
\bea
	f(w) &=& {3\over 4}\left\{{1+w\over w^3}\left[{2w(1+w)\over 1+2w} -
\ln(1+2w)\right] \right.\nonumber\\
&+& \left.{\ln(1+2w)\over 2w} - {1+3w\over (1+2w)^2}\right\}
\eea
The Thomson cross section is given by $\sigma_T =
(8\pi/3)\alpha\alpha'(\epsilon/m_\de)^2$.  
For the thermal average, we take
\be
	\langle\sigma v\rangle_{\gamma\to\gamma'} = 
	\int {d^{3\,}p_\de\over (2\pi)^3}
	\int {d^{3\,}p_{\gamma'}\over (2\pi)^3}
	\,f_\gamma\, f_\de\,
	\sigma(w)|\vec v_{\rm rel}|
\label{thermavg}
\ee
where $w=\sfrac12(s/m_\de^2-1)$ in terms of the Mandelstam invariant
$s=(E_\de +E_\gamma)^2 - (\vec p_\de + \vec p_\gamma)^2$ and 
$\vec v_{\rm rel} = \vec p_\de/E_\de - \vec p_\gamma/E_\gamma$.
For the process $\gamma\de\to\gamma'\de$, we allow for different
temperatures in the two sectors, so that the (normalized) distribution functions
have the dependences 
$f_\gamma= f_\gamma(p_\gamma,T_\gamma)$, 
$f_\de = f_\de(p_\de,T_d)$.   In contrast, for  
$\gamma'\de\to\gamma\de$, we take both initial particles to have
temperature $T_d = T_{\gamma'}$, since we assume that
$\gamma'\de\leftrightarrow\gamma'\de$ scattering is strong enough to
keep the dark particles in kinetic equilbrium for as long as there is
a signficant $\de$ population.  Carrying out the thermal average
in (\ref{thermavg}) numerically, we obtain the results shown in fig.\
\ref{fig:therm} for a range of visible to dark temperature ratios,
$\zeta = T/T_d$.  We find that $\langle\sigma v\rangle/\sigma_T$
can be approximated by the analytic form
\be
	{\langle\sigma v\rangle\over\sigma_T}
	\cong \left(\sfrac12\left(1 + \tanh(A_0\log_{10}x-A_1)
	\right)\right)^{A_2}
\label{svfit}
\ee
where $x=m_\de/T_d$ and the coefficients $A_i$ depend upon $\zeta$
as shown in table \ref{tab1}.  We can thus 
rapidly interpolate between the
$\zeta$ values of interest without having to repeat the integration
in (\ref{thermavg}).

\begin{table}[t]\centering
\begin{tabular}{|c|c|r|c|}\hline
$\zeta$ & $A_0$ & $A_1$ & $A_2$\\
\hline
0.1  &2.209   &0.041 &0.583 \\
0.25 &1.420  &$-$0.0056 &1.156\\
0.5  &1.099  &$-$0.092 &1.843 \\
1    &0.934 &$-$0.037 &2.265\\
2    &0.854  &0.189  &2.112\\
4    &0.817  &0.496  &1.773\\
10   &0.786  &0.878  &1.496\\
\hline
\end{tabular}
\caption{Coefficents for the analytic fit (\ref{svfit}) to 
the thermally averaged $\gamma'\de\leftrightarrow\gamma\de$ scattering cross section.}
\label{tab1}
\end{table}

For the simplified approach in which we estimate the freezeout
temperature $T_f$ via $\Gamma = n_\de\langle\sigma v\rangle = H$,
we take $\zeta = 1$.  In the more quantitative Boltzmann analysis,
the distinction enters into the source term $q_{\rm scatt}$ which
we take to be
\be
	q_{\rm scatt} = n_\de\left(\langle\sigma
v\rangle_{\gamma\to\gamma'}\rho_\gamma - 
\langle\sigma
v\rangle_{\gamma'\to\gamma}\rho_{\gamma'}\right)
\ee
where $\rho_i$ are the energy densities.  The cross
section $\langle\sigma
v\rangle_{\gamma\to\gamma'}$ depends upon $\zeta$ whereas
$\langle\sigma
v\rangle_{\gamma\to\gamma'}$ depends only upon the dark photon
temperature through $m_\de/T_d$ (and corresponds to the case $\zeta=1$).
When $q_{\rm scatt}$ is 
large compared to $H\rho_i$, it drives the two photon baths
toward equilibrium with each other.

\begin{figure*}[t]
\centerline{\includegraphics[width=\columnwidth]{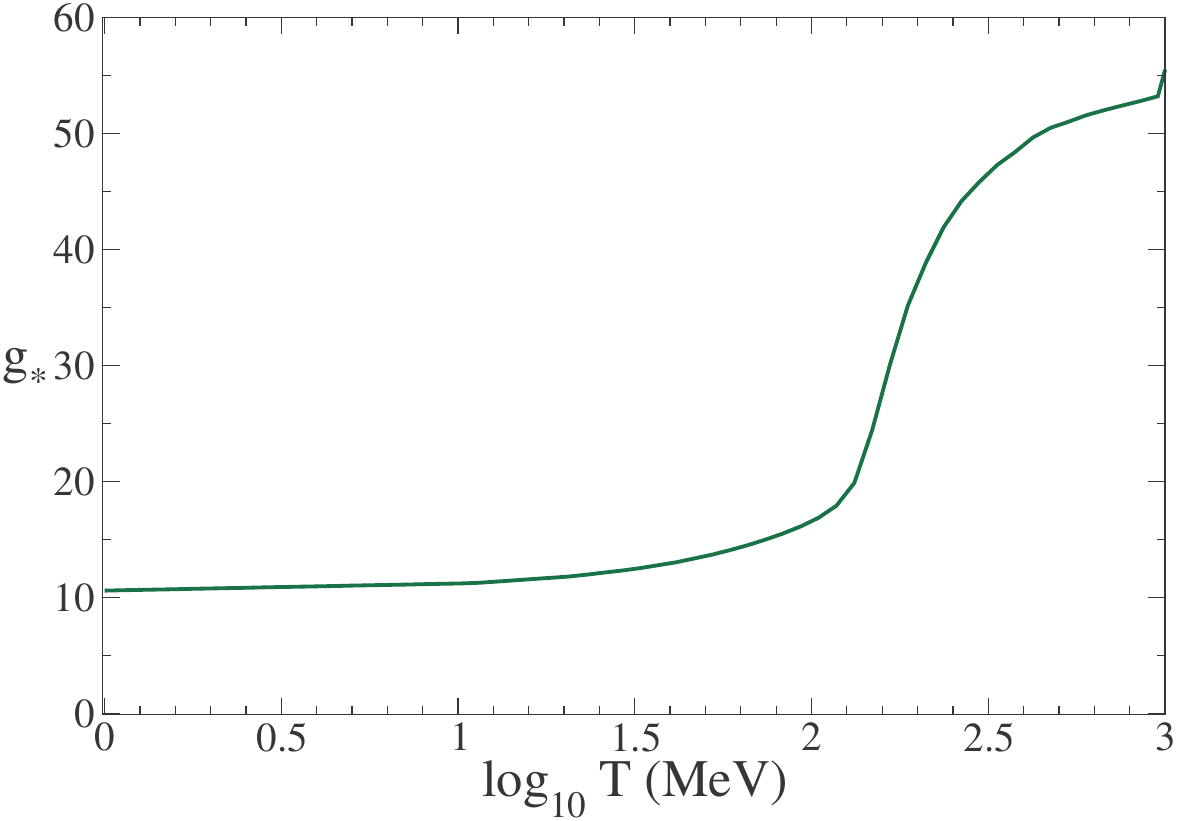}
\includegraphics[width=\columnwidth]{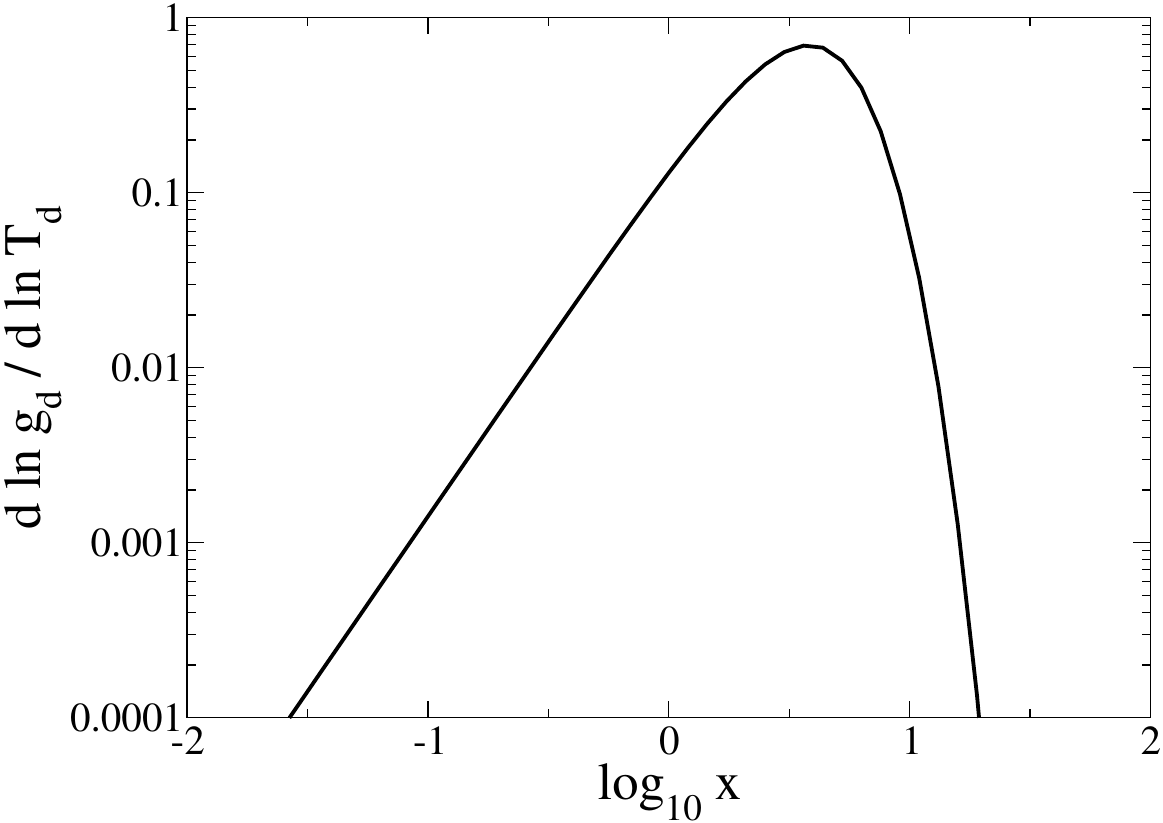}}
\caption{Left: Effective number of degrees of freedom $g_*$ versus
temperature, inferred from ref.\ \cite{Borsanyi:2010cj}.
Right: Dependence of $d\ln g_d/d\ln T_d$ on $x = m_\de/T_d$,
appearing in the source term $q_{\rm ann}$.
}
\label{fig:gstar}
\end{figure*}

The other source terms in the Boltzmann equations are given by
\bea
	q_{\rm\sss SM} &=& \sfrac43 H {\rho_\gamma {d\ln g_*\over
	d\ln T_\gamma}\over 1 + \frac13 {d\ln g_*\over
	d\ln T_\gamma}}\\
	q_{\rm ann} &=&  \sfrac43 H {\rho_{\gamma'} {d\ln g_d\over
	d\ln T_d}\over 1 + \frac13 {d\ln g_d\over
	d\ln T_d}}
\eea
where $g_d = 2 + g_\de$ is the effective number of degrees of 
freedom in the dark plasma. 
These are a straighforward consequence of entropy conservation
as the photons get heated by annihilation of heavier standard model
particles,  or the dark photons are heated by dark electron annihilation.
For reference we display $g_*(T)$
in fig.\ \ref{fig:gstar}(left).  It is found by adding the contributions
from photons and leptons to the QCD degrees of freedom determined
by ref.\ \cite{Borsanyi:2010cj}.  The dependence of 
$d\ln g_d/d\ln T_d$ on $x = m_\de/T_d$ is shown in fig.\ 
\ref{fig:gstar}(right).

\section{Pion scattering at low energy}
\label{chipt}

In this appendix we provide details of the elastic cross section
for dark pion scattering at low energy, derived from the chiral
Lagrangian (\ref{chiral_lag}).  Expanding to fourth order in the
pion field $\Pi$, the relevant interaction terms are
\be 
   \mathcal{L}_{4\pi} = \frac{ 1 }{3 F_\pi^2} 
\Tr\left[ \left( \Pi\, \overset\leftrightarrow{\partial^\mu} \Pi \right)
      \left( \Pi\, \overset\leftrightarrow{\partial_\mu} \Pi \right)  
+ m_\pi^2 \Pi^4 \right]
\label{eq:L4pi}
\ee
The matrix element of the process 
$ \pi^a + \pi^b \rightarrow \pi^c + \pi^d$ derived from 
\ref{eq:L4pi}) is
\bea
   -i\mathcal{M}
 &=& \left( \Tr \left[ T^a T^b T^c T^d \right] + ( b \leftrightarrow d)  \right) 
      \frac{2( 2 m_\pi^2 - t) } {  F_\pi^2 }+ \nonumber\\
   && \left( \Tr \left[ T^a T^c T^b T^d \right] + ( c \leftrightarrow d) \right) \frac{ 2( 2 m_\pi^2 - s
     )}  {  F_\pi^2 }+ \nonumber\\
   && \left( \Tr \left[ T^a T^c T^d T^b \right] + ( b \leftrightarrow c) \right) 
      \frac{2( 2 m_\pi^2 - u  ) }{  F_\pi^2 } \nonumber\\
\eea
In the case of SU(2), 
$   \Tr [ T^a T^b T^c T^d ]$ $ = \frac{1}{8} $ $( \delta_{ab} \delta_{cd} + $ 
$\delta_{ad}\delta_{bc} - \delta_{ac}\delta_{bd})$
 \cite{Weinberg:1966kf,Scherer:2002tk}. For general SU($N$), 
\be   
   \Tr [ T^a T^b T^c T^d ] = \frac{1}{4N}  \delta_{ab} \delta_{cd}  + \frac{1}{8} \left( d^{abe} + i f^{abe} \right)
      \left( d^{cde} + i f^{cde} \right)
\ee
and the isospin-averaged matrix element 
squared (using Mathematica) is
\begin{widetext}
\bea   
   | \mathcal{M} |^2  &=& \frac{1} { (N^2 -1)^2} \sum_{abcd} | \mathcal{M}_{abcd} |^2  \nonumber
   \\
   &=& \frac{ \left[  8  \left( 2 N^4 -25N^2+90 - 65 N^{-2}\right) m_\pi^4- \left(3N^4 - 37 N^2 + 132 -96N^{-2}\right) 
      \left( s~t + t~u + u~s \right) \right]} { 2 F_\pi^4 ( N^2 -1) }
\eea
Therefore, the cross section at center-of-mass momentum $p$ is
\be
   \sigma_{2\pi\rightarrow 2\pi} = \frac {\left[ \left( 2 N^4 -25N^2+90  - 65N^{-2}\right)  m_\pi^4+
 2\left(3N^4 - 37 N^2 + 132  -96N^{-2}\right)
      \left(  m_\pi^2 p^2 - \frac{5}{6}p^4 \right) \right]}{ 32 \pi F_\pi^4 ( N^2 -1) (m_\pi^2 +p^2)}
\ee
\end{widetext}

\section{Pion coannihilation to a light gauge boson}
\label{pi-ann}

The interaction (\ref{zprime}) does not respect the full SU(3)$_L\times$SU(3)$_R$
chiral flavor symmetry $\Sigma \to V^\dagger\Sigma U$, but it does
respect the diagonal subgroup $\Sigma \to V^\dagger\Sigma V$, requiring
the quark mass insertion that we have made explicit.  (The quark mass
matrix takes the place of the missing $\Sigma^\dagger$ field.)
Expanding (\ref{zprime}) to leading order in the pion fields gives
\be
	-4{\lambda_0\,m_q\over F_\pi^4}\,Z'_{\mu\nu}\, f_{abc}
	\pi^a\, \partial^\mu\pi^b\,\partial^\nu\pi^c
\ee
Defining $\tilde\lambda = 4{\lambda_0\,m_q/ F_\pi^4}$,
the matrix element for the process $\pi^a(p_1)+\pi^b(p_2)\to
\pi^c(p_3) + Z'(p_4)$ is (up to a phase)
\be
	2\tilde\lambda(\epsilon^\mu p_4^\nu 
-\epsilon^\nu p_4^\mu)f_{abc}\left[p_2^\mu p_3^\nu - 
	p_1^\mu p_3^\nu - p_1^\mu p_2^\nu \right]
\ee
where $\epsilon^\mu$ is the polarization vector of $Z'$.
The isospin-averaged, squared matrix element is given by
\bea
	\langle|{\cal M}|^2\rangle &=& 
-{9\tilde\lambda^2 N_f\over 4(N_f^2-1)}\left(st(s+t)-3 m_\pi^2 st +
	m_\pi^6\right)\nonumber\\
	&\cong& {135\,\tilde\lambda^2 N_f\over 4(N_f^2-1)}\,m_\pi^6\,v^4
\eea
where the second line gives the low-energy limit, with $v$ being the
velocity of the incoming particles in the c.m.\ frame.

\end{document}